\documentclass[twocolumn,11pt]{article}
\usepackage[margin=1.5cm]{geometry}
\usepackage{amsmath}
\usepackage{amssymb}
\usepackage{microtype}
\usepackage{abstract}
\usepackage[small,bf]{caption}
\usepackage{graphicx}
\usepackage{color}

\newcommand{\act}{\zeta\Delta\mu}
\newcommand{\actb}{\hat{\zeta}\Delta\mu}
\newcommand{\actt}{\tilde{\zeta}\Delta\mu}
\newcommand{\acttp}{\zeta^{\prime}\Delta\mu}
\newcommand{\divv}{\nabla \cdot}   
\newcommand{\curl}{\nabla \times}
\newcommand{\vectl}[1]{\boldsymbol{#1}}  
\newcommand{\vects}[1]{\boldsymbol{#1}}    
\newcommand{\tenss}[1]{\mathsf{\boldsymbol{#1}}} 
\newcommand{\curlb}[1]{\left\{ #1 \right\}}
\newcommand{\sqb}[1]{\left[ #1 \right]}
\newcommand{\rndb}[1]{\left( #1 \right)}
\newcommand{\norm}[1]{\left| #1 \right|}
\newcommand{\diff}[2]{\frac{\rm d#1}{\rm d#2}}
\newcommand{\pdiff}[2]{\frac{\partial#1}{\partial#2}}
\newcommand{\si}[1]{\,{\rm{#1}}}
\newcommand{\mcrm}{\,\mu{\rm{m}}}

\begin{document}

\twocolumn[
\begin{center}
\textbf{\Large Active polar fluid flow in finite droplets}\\
\vspace{7mm}
\underline{C. A. Whitfield$^1$}, D. Marenduzzo$^2$, R. Voituriez$^{3,4}$ and R. J. Hawkins$^1$\\
\vspace{7mm}
\emph{1. Department of Physics and Astronomy, University of Sheffield, Sheffield, United Kingdom}\\
\emph{$^2$ SUPA, School of Physics and Astronomy, University of Edinburgh, Mayfield Road, Edinburgh EH9 3JZ, United Kingdom}\\
\emph{$^3$ Laboratoire de Physique Th\'{e}orique et Mati\`{e}re Condens\'{e}e, UMR 7600, Universit\'{e} Pierre et Marie Curie/CNRS, Paris, France}\\
\emph{$^4$ Laboratoire Jean Perrin, FRE 3231 CNRS /UPMC, 4 Place Jussieu, F-75255 Paris Cedex}
\vspace{7mm}
\end{center}
\vspace{-8mm}
  \begin{onecolabstract} 
We present a continuum level analytical model of a droplet of active contractile fluid consisting of filaments and motors. We calculate the steady state flows that result from a splayed polarisation of the filaments. We account for interaction with the external medium by imposing a viscous friction at the fixed droplet boundary. We then show that the droplet has non-zero force dipole and quadrupole moments, the latter of which is essential for self-propelled motion of the droplet at low Reynolds' number. Therefore, this calculation describes a simple mechanism for the motility of a droplet of active contractile fluid embedded in a three-dimensional environment, which is relevant to cell migration in confinement (for example, embedded within a gel or tissue). Our analytical results predict how the system depends on various parameters such as the effective friction coefficient, the phenomenological activity parameter and the splay of the imposed polarisation.\\
  \end{onecolabstract}
]

\section{Introduction}
\label{sec:intro}

The study of active matter, \emph{i.e.} collections of particles that are driven out of equilibrium individually by internal energy, has successfully been applied to various biological and physical systems \cite{Ramaswamy2010}, such as flocks of animals or bacteria \cite{Toner1998} or vibrated granular rods \cite{Narayan2007}. The hydrodynamic model of active gels developed by Kruse \emph{et al.} \cite{Kruse2000,Kruse2004,Kruse2005} considers the case where the active matter in question is a viscoelastic gel comprising of polar filaments that are pulled by motor molecules generating active stresses in the gel. This can be used as a model of the actin cytoskeleton, where the active stresses generated are contractile (to model the interaction between myosin II motor proteins and F-actin), as discussed in \cite{Joanny2009}. In this way, the theoretical model is analogous to \emph{in-vitro} experiments on reconstituted cytoskeletal networks, which probe the fundamentals of cellular mechanics by isolating specific components of the cytoskeleton and observing the interactions and behaviour \cite{Mizuno2007,Mackintosh2010,Silva2011,Kohler2011,Kohler2012}.\\

In this paper, we consider a motility mechanism that arises in a droplet of active gel due solely to active contractile stress from the interaction of myosin II and F-actin. This non-equilibrium activity is fuelled by energy released in the binding of Adenotriphosphate (ATP) to myosin II and its hydrolysis into Adenodiphosphate (ADP) by the motor proteins as they transiently attach and exert forces on adjacent filaments. Thus, as in \cite{Kruse2004} and \cite{Voituriez2005}, we ignore the self-polymerisation of the actin filaments so that we can isolate the effects of the contractile stresses generated by the actomyosin network only. We apply a viscous friction condition at the boundary to address the issue of cell motility in three-dimensional (3D) confinement (such as tissue-like environments), the importance of which is being increasingly recognised \cite{Charras2008,Hawkins2009,Konstantopoulos2013}, particularly due to new experimental techniques that enable its direct observation \cite{Poincloux2011,Weber2013}.\\ 

We compare our results to the lattice Boltzmann fluid simulations of an active droplet by Tjhung \emph{et al.} \cite{Tjhung2012}, which shows that a droplet of active gel immersed in a Newtonian fluid will display spontaneous symmetry breaking when the intensity of motor activity is above a certain threshold. This symmetry breaking causes the droplet to reach a motile steady state, similar to that discussed in this paper. The case we present considers a system where the symmetry is already broken by a splayed polarisation of the filaments, which we justify in sect. \ref{sec:model}. Imposing the polarisation field allows us to model this system analytically, providing greater insight into the important factors behind this motile steady state. Also the droplet modelled here interacts with the external medium via a linear viscous friction, whereas the simulations in \cite{Tjhung2012} use a two-phase model with periodic boundary conditions, where the passive viscous properties of the two phases of fluid are identical. Therefore, { we are able to predict how the steady state flow in the droplet depends on the effective friction with the external medium.}\\   

{ There are various examples of studies that model the dynamics of active droplets in the contrasting situation where the droplet is adhered to a surface. For example, Blanch-Mercader and Casademunt show analytically in \cite{Blanch-Mercader2013} that an actin lamellar fragment is unstable to perturbations in shape when polymerisation forces act at the membrane. Similarly, in \cite{Shao2010} and \cite{Ziebert2011} the authors use phase field based approaches to model the crawling motility of an actin gel driven by polymerisation at the leading edge. These studies all suggest an underlying mechanical mechanism for the observed shapes of crawling cells and cell fragments. In this paper we investigate a motile steady state of an active droplet in suspension driven only by actomyosin contraction and explain how this results in the steady state observed in simulations.}\\

We first present the two-dimensional (2D) version of the calculation to simplify both the mathematics and graphical representation of the system. This allows for comparison between the analytical results and new lattice Boltzmann simulations of an active droplet that use the source code introduced in \cite{Tjhung2012}. In {sect. \ref{sec:sphere} and appendix \ref{app:sphere}} we show how the analytical case generalises to 3D.
 
\section{Model}
\label{sec:model}

We begin with a circular droplet of active gel of radius $R$ and assume that the boundary remains fixed. This assumption is valid for a droplet with high cortical tension, such that any active pressure gradients in the droplet would have a negligibly small effect on the boundary shape. This is the case for \emph{in-vitro} emulsion droplets of cytoskeletal filaments \cite{Sanchez2012} as the droplets in these cases do not visibly deform from spherical even in the presence of high activity. MDA-MB-231 breast tumour cells in matrigel provide an { \emph{in-vivo} example of such a system}, as they can migrate through matrigel while maintaining a nearly spherical shape \cite{Poincloux2011}. { A thorough calculation of the resulting droplet deformation in such a system would require numerical simulation and we leave this interesting investigation to future work.}\\

We define the polarisation $\vectl{p}$ as the average alignment direction of the `barbed' ends of the actin filaments at a given point in the droplet and assume that, on average, the filaments in the active fluid are highly ordered and thus the gel is far from the isotropic phase. This means that the magnitude of the polarisation can be defined as $\norm{\vectl{p}} = 1$ without loss of generality. { This assumption is valid for networks of high concentration, since the nematic order of actin networks increases with filament concentration \cite{Coppin1992}. Variations in the order parameter ($\norm{\vectl{p}}$) can be included as in \cite{Ziebert2011} where numerical methods are used to simulate a crawling keratocyte fragment.}

\subsection{Constitutive equations}
\label{sec:equations}

To model the dynamics of the internal active gel, we use the coarse grained hydrodynamic approach outlined by Kruse \emph{et al.} \cite{Kruse2000,Kruse2004,Kruse2005}. We take the long time limit ($t \gg \tau$ where $\tau$ is the relaxation time of the gel), which models the active gel as a viscous fluid. { Then, the total stress can be written as follows,
\begin{align}
\label{totstress} \sigma_{\alpha \beta}^{\mathrm{tot}} = \sigma_{\alpha\beta}^{\mathrm{visc}} + \sigma_{\alpha\beta}^{\mathrm{dist}} + \sigma_{\alpha\beta}^{\mathrm{act}}\si{.}
\end{align}
The first contribution is the viscous stress which is proportional to the rate of strain in the fluid,
\begin{align}
\label{viscstress} \sigma_{\alpha\beta}^{\mathrm{visc}} = 2\eta u_{\alpha\beta} = \eta\rndb{\partial_\alpha v_\beta + \partial_\beta v_\alpha}  \si{,}
\end{align}
where $\eta$ is the shear viscosity and $\vectl{v}$ is the fluid velocity. Next, the stress caused by distortions in the nematic alignment of the filaments,
\begin{align}
\label{diststress} \sigma_{\alpha\beta}^{\mathrm{dist}} = \frac{\nu}{2} \rndb{p_\alpha h_\beta + p_\beta h_\alpha} + \frac{1}{2}\rndb{p_\alpha h_\beta - p_\beta h_\alpha} + \sigma_{\alpha\beta}^{\mathrm{e}} \si{,}
\end{align}
where the molecular field $\vectl{h} = -\rndb{\delta F_d}/\rndb{\delta \vectl{p}}$ and $F_d$ is the distortion free energy for a network of polar filaments in the passive regime with $\norm{\vectl{p}}=1$. This is given by}
\begin{align}
F_d = \int f_0 \si{d}\vectl{r} = \int \frac{1}{2}\left\{K_1 \rndb{\divv \vectl{p}}^2 + K_2 \sqb{\vectl{p} \cdot \rndb{\curl \vectl{p}}}^2 \right.\notag \\
\label{freeE} \left. + K_3 \sqb{\vectl{p} \times \rndb{\curl \vectl{p}}}^2 { + k\rndb{\divv{\vectl{p}}}} \right\}\si{d}\vectl{r} \si{,}
\end{align} 
where $K_1$, $K_2$ and $K_3$ are the elastic coefficients for the terms corresponding to splay, twist and bend deformations respectively \cite{Gennes1993}. { The final term, proportional to $k$, is the spontaneous splay, which is allowed by symmetry because of the polar nature of the filaments. However, this term is only relevant when the liquid crystal is near to the isotropic to polar transition \cite{Pleiner1989} and since we take $\norm{\vectl{p}}=1$ (as discussed in sect. \ref{sec:model}) we ignore it in our analysis.} { The distortion stress from eq. \eqref{diststress} depends also on $\nu$, which is a dimensionless constant that relates to the coupling between the polarisation field and the flow (negative for rod-like particles; the value and sign of $\nu$ however have no qualitative bearings on our results). The second term in eq. \eqref{diststress} is the anti-symmetric component of the distortion stress and { the final term is the contribution from the Ericksen stress (as discussed for active gels in \cite{Furthauer2012}),
\begin{align}
\label{estress} { \sigma_{\alpha\beta}^{\mathrm{e}} = f_0\delta_{\alpha\beta} - \pdiff{f_0}{\rndb{\partial_\beta p_\gamma}} \partial_\alpha p_\gamma} \si{.}
\end{align}}

The final contribution to eq. \eqref{totstress} is from the active stress, which we assume to be traceless as we assume conservation of the droplet volume,
\begin{align}
\sigma_{\alpha\beta}^{\mathrm{act}} = - \act \rndb{p_\alpha p_\beta - \frac{\delta_{\alpha\beta}}{d}} \si{.}
\end{align}
The parameter $\zeta$ is a phenomenological measure of the activity strength, $\Delta\mu$ is the difference in chemical potential between ATP and ADP, and $d$ is the dimensionality of the system. Actomyosin networks create contractile stresses \cite{Bendix2008}, and this corresponds to $\zeta < 0$ \cite{Kruse2004}. As in \cite{Kruse2004}, only active terms linearly proportional to $\zeta$ are considered, describing the case where the system is linearly out of equilibrium.}\\

{ When the system is in steady state, the total stress will satisfy the force balance eq.:
\begin{align}
\label{fbgen} \partial_\alpha \rndb{\sigma_{\alpha \beta}^{\mathrm{ tot}} - P\delta_{\alpha \beta}} = 0 \si{,}
\end{align} 
where $P$ is the internal hydrodynamic pressure}. Equation \eqref{fbgen} has no inertial terms since we work in the low Reynolds number limit, due to the small length scale and velocities that are involved at the cellular level. { As we take the fluid limit, we also assume incompressibility, which is imposed by the condition,}
\begin{align}
\label{incgen} \divv \vectl{v} = 0 \si{.}
\end{align}

{ Finally, the dynamic eq. for the polarisation field $\vectl{p}$ in an incompressible active fluid is given by,
\begin{align}
\pdiff{p_\alpha}{t} =& -\rndb{v_\gamma\partial_\gamma}p_\alpha - \omega_{\alpha\beta}p_\beta - \nu u_{\alpha\beta}p_\beta  \notag \\
\label{dpdt} &+ \frac{1}{\gamma}h_\alpha + \lambda p_\alpha\Delta\mu \si{,}
\end{align}
where $\omega_{\alpha \beta} = \rndb{\partial_\alpha v_\beta - \partial_\beta v_\alpha}/2$ is the vorticity tensor, $\gamma$ is the rotational viscosity and $\lambda$ is a phenomenological active parameter.

\subsection{Imposed filament polarisation}
\label{sec:polarisation}

 In previous analytical studies, it has been shown that eqs. \eqref{totstress} and \eqref{dpdt} predict that contractile active fluids are unstable to splay defects.} Firstly, in studies of infinite films \cite{Voituriez2006} and quasi one-dimensional active gels above the Friedrick's transition \cite{Voituriez2005}, a finite polarisation gradient and spontaneously flowing state is calculated, indicating that such a splayed polarisation field is a natural state in the active contractile phase. This generic splay instability in active contractile systems of filaments and motors is explained qualitatively by \cite{Ramaswamy2010} as due to long wavelength splay fluctuations, which perturb the balance of flow by pulling fluid along the axes of the filaments, creating a shear. This shear results in an amplification of the splay fluctuation, causing a feedback loop.\\

\begin{figure}[h]
\vspace{-3mm}
\centering
\includegraphics[width=0.45\textwidth]{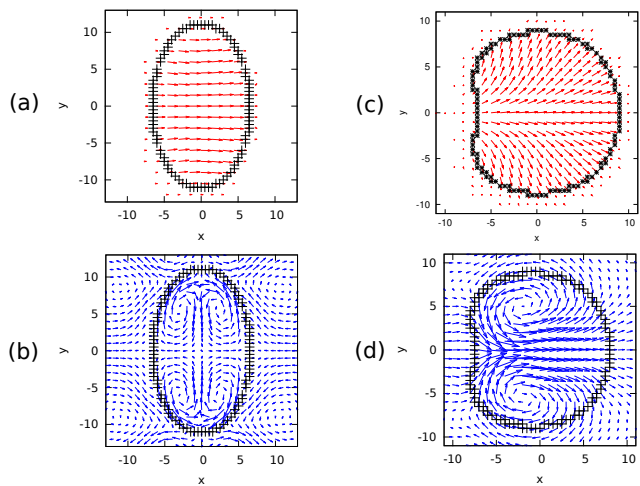}
\caption{{Lattice Boltzmann simulations of an active contractile droplet with the following parameter values in simulation units: $K=0.04$, $\nu = -1.1$, $\nu = 5/3$, $\gamma = 1$ and $\act = -0.005$. Also, a higher value of surface tension was used than in \cite{Tjhung2012}: $k_\phi = 0.3$.} \textbf{(a)} Polarisation field before symmetry breaking. \textbf{(b)} Velocity profile before symmetry breaking. \textbf{(c)} Splayed polarisation field in motile steady state \textbf{(d)} Velocity profile of the motile steady state in the droplet reference frame.}   
\label{fig:vsim}
\vspace{-3mm}
\end{figure}

{ Furthermore, using lattice Boltzmann simulations of the full dynamic equations (with the same source code as in \cite{Tjhung2012}), we find that in 2D a droplet of contractile active fluid (with no filament self-advection) is unstable to splay defects in the polarisation. The simulations use a phase field to define the active and passive phases of the fluid, as outlined in appendix \ref{app:sims}. The results in 2D show that, above a certain critical value of $\zeta$, the droplet reaches a splayed steady state, which is plotted in fig. \ref{fig:vsim}(c).} { This splayed steady state is stable despite the fact that it results in vortices in the flow (fig. \ref{fig:vsim}(d)).} \\

In the case of confinement in a droplet, boundary effects become important and these can promote splay in the filament polarisation depending on the specific anchoring conditions at the boundary. For example, cells with broken symmetry often display a higher concentration of actin nucleators at the leading edge, which polarises the actin there. { However, we do not consider any anchoring at the boundaries in the lattice Boltzmann simulations presented in fig. \ref{fig:vsim}.}\\

{ Based on this evidence,} we investigate the effects of polarisation splay in an active fluid droplet by imposing a splayed polarisation field. {We then treat $\vectl{p}$ as fixed in time, modelling only the steady state flow.} { Additionally, we assume that $\zeta$ is constant for this calculation, which can be interpreted as assuming that there is a uniform density of ATP and myosin II throughout the active gel.} { This assumption of constant $\zeta$ means that we potentially miss some interesting effects due to gradients in the activity. On hydrodynamic time scales the motors would be advected by the flow which could lead to a feedback mechanism (\emph{e.g.} as calculated in \cite{Hawkins2011} for a compressible cortical layer of active gel), but we would not expect this effect to occur in incompressible systems. On shorter time scales it is likely that a splay in the polarisation would lead to an inhomogeneous distribution of the myosin II motors that mediate contraction, due to these motors `walking' in the direction of the filament barbed ends \cite{Alberts2008}. However we do not consider this here, but instead focus on the effects of contraction in such a splayed state.}\\ 

\begin{figure}[h]
\centering
\includegraphics[width=0.4\textwidth]{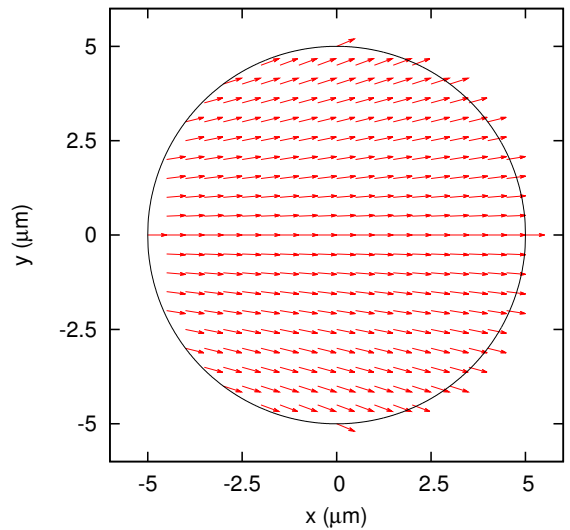}
\centering
\caption{Vector plot of the polarisation field $\vectl{p}$ of eq. \eqref{pl} with length scale $l=20\,\mu\rm{m}$ imposed on a circular droplet of radius $R = 5\,\mu\rm{m}$.}
\label{fig:pol}
\end{figure}

 As we are working in the $\norm{\vectl{p}} = 1$ limit, we can assume without loss of generality that $p_x = \cos(\psi)$ and $p_y = \sin(\psi)$ where $\psi$ is the angle between the filament polarisation and the $x$-axis. Then, we can arbitrarily choose a direction for the splay to occur, in this case we specify that the polarisation should splay outwards from the $x$-axis. {The simplest example of this is if $\psi$ is an anti-symmetric function of $y$ only. Thus, if we take $\psi$ to be linear in $y$ then we find a polarisation that is qualitatively similar to that in fig. \ref{fig:vsim}(c), and hence we impose the polarisation:
\begin{align}
\label{pl} \vectl{p} = \sqb{\cos\rndb{\frac{\pi y}{2l}},\sin\rndb{\frac{\pi y}{2l}}} \si{,}
\end{align}
where $l$ defines the length scale of the variation over $y$ (see fig. \ref{fig:pol})}. The relative amount of splay and bend in the polarisation can be calculated by finding the { ratio of the magnitudes of the `splay' and `bend' terms in the distortion free energy (from eq. \eqref{freeE}). Taking the one constant approximation $K_1 = K_2 = K_3 = K$, this ratio is:
\begin{align}
\label{splayvbend}  \frac{\rndb{\divv \vectl{p}}^2}{\sqb{\vectl{p} \times \rndb{\curl \vectl{p}}}^2} = \cot^2\rndb{\frac{\pi y}{2l}} \si{.}
\end{align}
This shows that splay is the dominant distortion everywhere in the droplet when $l>2R$ (note that the `twist' term, $\sqb{\vectl{p} \cdot \rndb{\curl \vectl{p}}}^2$, is always zero in the 2D case, only contributing in 3D) and so we take this as a minimum value for $l$. In this limit, we can define the parameter $c_s = R/l$ as a measure of the magnitude of the splay in the droplet, the maximum value of which is $\rndb{\divv \vectl{p}}^2 = (\pi c_s/(2R))^2$ at $y=0$.}

Substituting eq. \eqref{pl} into eq. \eqref{totstress}, we find expressions for the total stress tensor $\tenss{\sigma}^{\mathrm{ tot}}$. Using eqs. \eqref{fbgen} and \eqref{incgen} we arrive at the general steady state equations of motion for the system:
\begin{align}
\label{pde1} \eta {\nabla}^2v_x(x,y) &= \frac{\actt\pi}{2l}\cos{\left(\frac{\pi y}{l}\right)} + \partial_x P(x,y) \si{,} \\
\label{pde2} \eta {\nabla}^2v_y(x,y) &= \frac{\actt\pi}{2l}\sin{\left(\frac{\pi y}{l}\right)} + \partial_y P(x,y) \si{,} \\
\label{pde3} 0 &= \vphantom{\frac{\actt\pi}{2l}}\partial_x v_x(x,y) + \partial_y v_y(x,y)\si{,}
\end{align}
where $\actt = \act + \nu K\pi^2/\rndb{4l^2}$. { To acquire analytical solutions, we expand eqs. \eqref{pde1} and \eqref{pde2} in terms of Taylor series up to quadratic order in $y/l$,}
\begin{align}
\eta {\nabla}^2v_x(x,y) &= \frac{\actt\pi}{2l}\sqb{1-\frac{\pi^2 y^2}{2l^2} + \mathcal{O}\rndb{\frac{y^3}{l^3}}} \notag \\
\label{pde1s} & + \, \partial_x P(x,y) \vphantom{a_{b_{c_{d_{e_{f}}}}}} \si{,} \\ 
\label{pde2s} \eta {\nabla}^2v_y(x,y) &= \frac{\actt\pi}{2l}\sqb{\frac{\pi y}{2l} + \mathcal{O}\rndb{\frac{y^3}{l^3}}} + \partial_y P(x,y) \si{.}
\end{align}
{This restricts our analysis to the limit $c_s \ll 1$, the regime where bend deformations are negligible relative to splay. To account for larger values of $c_s$, we are able to solve for Taylor series of arbitrarily high order, as discussed in appendix \ref{app:sols}. In this large $l$ limit, we can assume that the distortion contributions to the stress and flow equations should small compared to the active contribution ($\norm{K/l^2}<\norm{\act}$) and hence $\tilde{\zeta} < 0$ regardless of the value of $\nu$.}

\subsection{Boundary conditions}
\label{sec:boundaries}

We confine the solutions to a fixed circular droplet with the following boundary conditions:
\begin{align}
\label{circbc1} v_r(R,\theta) &= 0 \si{,} \\
\label{circbc2} \sigma_{r\theta}(R,\theta) &= -\xi v_\theta(R,\theta) \si{,}
\end{align} 
where $r$ and $\theta$ are standard polar coordinates and $R$ is the droplet radius. Equation \eqref{circbc1} ensures that there is no fluid entering or leaving the droplet and eq. \eqref{circbc2} applies an effective viscous friction at the boundary (with friction coefficient $\xi$). This friction condition is general as it infers little about the external medium, only that it will create some linear resistance to flow at the interface. If the droplet is embedded in a solid, then $\xi$ will determine the slip between the fluid and the boundary. Alternatively, if the external medium is a viscous fluid, and we assume non-slip between the internal and external fluid, then the friction coefficient $\xi$ will be related to the viscosity of the external fluid. Therefore we call $\xi$ the \emph{effective} friction coefficient. It is important to note that in a cell the conditions at the boundary would be more complicated, depending also on the adhesion between the membrane and the surrounding  environment, the elastic/viscoelastic properties of the external medium and the active processes of the membrane itself. { Including these effects would require more general boundary conditions that could be space and/or time dependant and in general would invalidate the assumption of a fixed boundary}.\\

The boundary conditions eqs. \eqref{circbc1} and \eqref{circbc2} are distinctly different from those used in lattice Boltzmann simulations of an active contractile droplet (as used to obtain fig. \ref{fig:vsim}; and introduced in \cite{Tjhung2012}). In these simulations, a phase field parameter is used to discern between the active and passive phases of the fluid, and this parameter is advected with the fluid velocity. Therefore the shape of the droplet is coupled to the flow, however the boundary between the two phases is diffuse and is defined by the gradient of the phase field. This also means that the external fluid in the simulations is assumed to have the same passive properties as the droplet. 

\section{Results and analysis}
\label{sec:results}

In general, we can solve eqs. \eqref{pde3}, \eqref{pde1s}, and \eqref{pde2s} by assuming power series solutions for $v_x$, $v_y$ and $P$:
\begin{align}
\label{vxpow} v_x(x,y) &= \sum\limits_{n=0}^{\infty}\sum\limits_{m=0}^{\infty}{a_{m,n}x^my^{n}}  \si{,}\\
\label{vypow} v_y(x,y) &= \sum\limits_{n=0}^{\infty}\sum\limits_{m=0}^{\infty}{b_{m,n}x^my^{n}}  \si{,}\\
\label{Ppow} P(x,y) &= \sum\limits_{n=0}^{\infty}\sum\limits_{m=0}^{\infty}{c_{m,n}x^my^{n}}  \si{.}
\end{align}
The coefficients $a_{m,n}$, $b_{m,n}$ and $c_{m,n}$ are arbitrary constants to be determined by the governing equations and boundary conditions. However, first we can impose that the solutions will be symmetric about the $x$ axis because the governing equations and boundary conditions have this symmetry. This leads to $a_{m,2n+1} = b_{m,2n} = c_{m,2n+1} = 0$ for all integer values of $m$ and $n$.\\

To find the final solution for a finite circular droplet, we can substitute the solutions eqs. \eqref{vxpow}, \eqref{vypow}, and \eqref{Ppow} into the approximated equations of motion; eqs. \eqref{pde3}, \eqref{pde1s}, and \eqref{pde2s} and boundary conditions; eqs. \eqref{circbc1} and \eqref{circbc2}. This gives an infinite number of simultaneous equations, but (due to the approximation made) one finds that $a_{m,n} = 0$, $b_{m,n} = 0$ and $c_{m,n} = 0$, if $m+n>6$, and so the series becomes finite. For more details see appendix \ref{app:sols}.

\subsection{Complete solutions}
\label{sec:sols}

\begin{figure*}
\centering
\includegraphics[width=\textwidth]{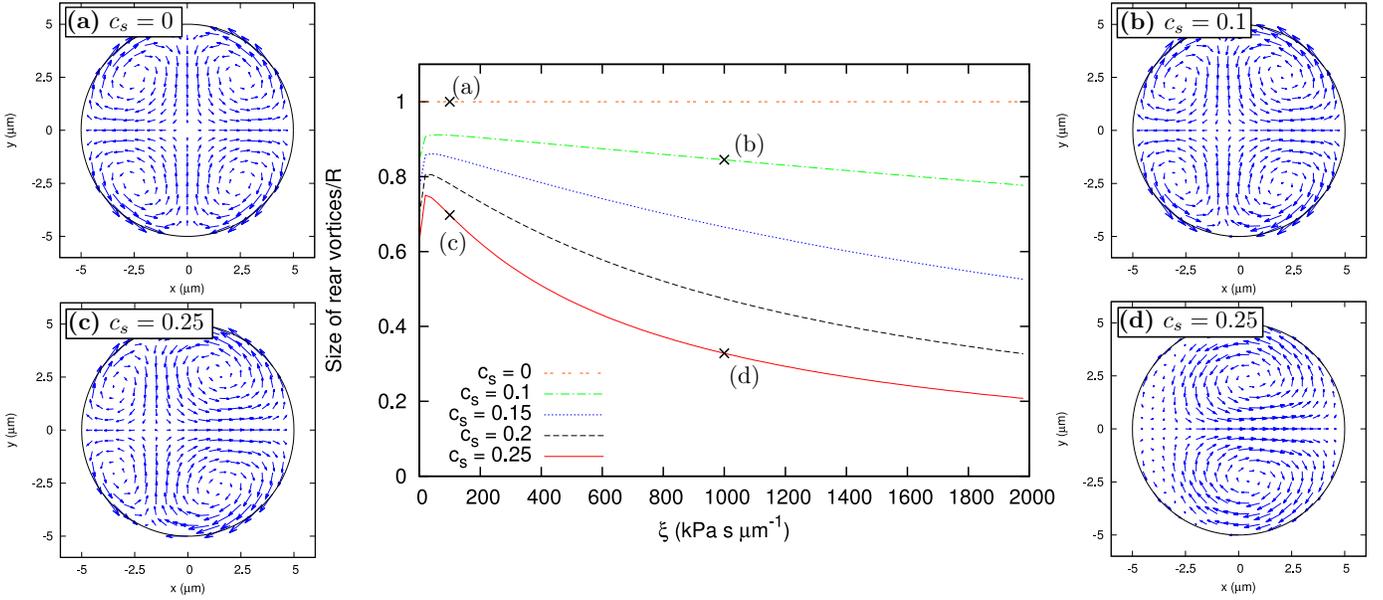}
\caption{Graph of the size of the two rear vortices in the droplet against the effective friction coefficient $\xi$, plotted for various values of the splay parameter $c_s$. Other parameter values are $\actt = -1\si{kPa}$ \cite{Kruse2006}, $K = 1\si{kPa\mcrm^2}$, $R = 5 \mcrm$, and $\eta = 10\si{kPa}\si{s}$ \cite{Wottawah2005}. Corresponding flow profiles at the points labelled \textbf{(a)}, \textbf{(b)}, \textbf{(c)}, and \textbf{(d)} are also shown. For these, the velocity arrow lengths are scaled independently for visibility by a scale factor $S$ such that a velocity magnitude of $v$ corresponds to an arrow length of $Sv$. The values of $S$ for each plot are: \textbf{(a)} and \textbf{(b)} $S = 300$, \textbf{(c)} $S = 3000$, and \textbf{(d)} $S = 1200$.}
\label{fig:tpflow}
\end{figure*}

\begin{figure}[h]
\centering
\includegraphics[width=0.45\textwidth]{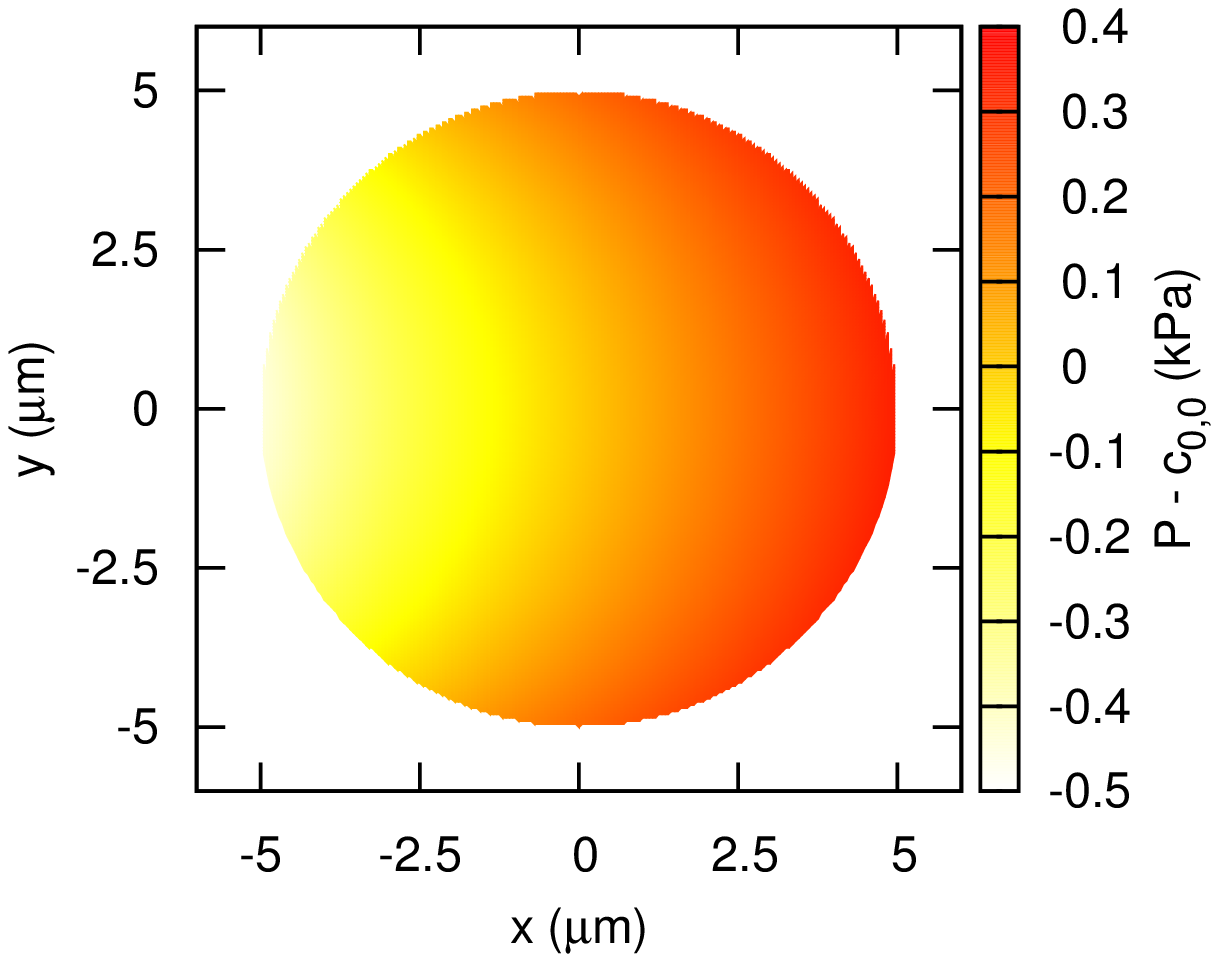}
\caption{Relative pressure inside the droplet for $\xi = 100 \si{kPa}\si{s}\mcrm^{-1}$. Other parameter values are $\actt = -1\si{kPa}$, $K=1\si{kPa}\mcrm^2$, $R = 5 \mcrm$, $\eta = 10\si{kPa}\si{s}$, and $c_s = 0.25$.}
\label{fig:press}
\end{figure}

The full 2D solutions are given in appendix \ref{app:sols} by eqs. \eqref{vxfull}, \eqref{vyfull} and \eqref{Pfull} and in this section they are presented graphically in figures \ref{fig:tpflow} and \ref{fig:press}.\\

Figure \ref{fig:tpflow} shows profiles of the velocity $\vectl{v}$ for different values of the effective friction coefficient $\xi$ and the splay parameter $c_s$. The velocity profile generally has 2 pairs of opposing vortices for lower values of $\xi$ and $c_s$, and as $\xi$ or $c_s$ is increased (the upper limit of $c_s$ is bound by the second order approximation in $r/l$) the front pair of vortices occupy more of the droplet. In the limit of zero splay, $c_s \rightarrow 0$ for fixed $R$ (or equivalently $l \rightarrow \infty$), the vortices become completely symmetric and the solutions are (fig. \ref{fig:tpflow}(a)):
\begin{align}
\label{vxnsp} v_{x} &= -\frac{\actt x}{R^2(\xi R + 4 \eta)}\rndb{R^2-x^2-3y^2} \si{,} \\
\label{vynsp} v_{y} &= \frac{\actt y}{R^2(\xi R + 4 \eta)}\rndb{R^2-3x^2-y^2} \si{,} \\
\label{Pnsp} P &=\, c_{0,0} + \frac{3\actt\eta}{R^2(\xi R + 4 \eta)}\rndb{x^2 - y^2} \si{.}
\end{align}
The source of the flow in this case is the remaining active terms in the boundary condition eq. \eqref{circbc2}, since in this limit there are no active terms present in the force balance eqs. \eqref{pde1} and \eqref{pde2}. This shows that even {when the polarisation field is completely aligned, the confinement of the active fluid to a droplet results in a flow (see fig. \ref{fig:tpflow}(a)), which is not found in studies of bulk active fluids. This observation is in agreement with the} lattice Boltzmann simulations of a droplet prior to symmetry breaking, which maintains an (approximately) aligned polarisation field (figures \ref{fig:vsim}(a) and (b)). Therefore the splay directly imposes the preferred direction in the flow, and hence the asymmetry of the vortices increases with $c_s$. The rear vortices only completely disappear mathematically in the infinite friction limit $\xi \rightarrow \infty$, because this destroys the boundary effects that induce the symmetric part of the flow.\\

The hydrodynamic pressure $P$ in eq. \eqref{Pnsp} still contains the undetermined constant $c_{0,0}$. This is the average pressure inside the droplet, and can be calculated by considering the forces that act in the direction normal to the droplet boundary. We assume that the net radial force will be zero, as we are considering the steady state solutions, which gives the condition:
\begin{align}
\label{Fbbc} \hat{\vectl{r}} \cdot \oint \sqb{\tenss{\sigma} - \rndb{P - P_{ext} - \frac{2\gamma}{R}}\hat{\mathbb{I}}} \cdot \rm{d}\vectl{s}\rndb{\theta}\, \bigg|_{r=R} = 0 \si{,}
\end{align}
where $\vectl{s}(\theta) = R\hat{\vects{\theta}}$ is the vector representation of the boundary curve at $r=R$, $\gamma$ is the surface tension of the droplet, and $P_{ext}$ is the pressure in the external medium (assumed to be constant). Solving eq. \eqref{Fbbc} for $c_{0,0}$ gives
\begin{align}
 \label{c00} c_{0,0} = \frac{2\gamma}{R} + P_{ext} + \frac{\actt\pi^2c_s^2}{16} \si{.}
\end{align}

Figure \ref{fig:press} plots the pressure difference inside the droplet for the same parameters as used in fig. \ref{fig:tpflow}(b). It shows that the variation in pressure across the boundary is approximately $1\si{kPa}$ with these estimated values. This means that, for our fixed boundary approximation to be valid, $2\gamma/R \gg 1 \si{kPa}$ meaning that (for these estimated values) $\gamma \gg 2.5 \si{kPa}\mcrm$. For comparison, experimental evidence suggests that the effective cell membrane tension (the combination of the bare membrane tension and the cortex tension) is $\gamma \approx 0.3\si{kPa}\mcrm$ \cite{Thoumine1999,Dai1999}. Thus, the high surface tension limit assumed in this calculation is not valid for the majority of cells. However, experiments on spontaneously moving active droplets consisting of microtubule filaments and motors show that droplets maintain near perfect spherical symmetry \cite{Sanchez2012}. In that case the active matter is confined to water droplets in oil so the interfacial tension is much larger.

\subsection{Analysis of solutions}
\label{sec:analysis}

The solutions simplify greatly in the infinite friction limit $\xi \rightarrow \infty$, which is equivalent to applying the non-slip boundary condition, $v_{\theta} = 0$ at $r=R$. Therefore, in this section we use these solutions to clarify the analysis and to keep the resulting equations brief. This analysis does generalise to the finite friction case, and we show how the results depend on the friction graphically. In the limit $\xi \rightarrow \infty$, the solutions become:
\begin{align}
\label{vxns}v_{x} &= \frac{\actt \pi^3 c_s^3}{384 \eta R^3}\rndb{R^2-x^2-y^2}\rndb{x^2+5y^2-R^2} \si{,} \\
\label{vyns} v_{y} &= -\frac{\actt \pi^3 c_s^3}{96 \eta R^3}xy\rndb{R^2-x^2-y^2} \si{,}\\
P &= c_{0,0} -\frac{\actt \pi c_s}{2R}\left[x + \frac{\pi c_s}{2R}\rndb{y^2}\right. \notag \\
\label{Pns} &+ \left.\frac{\pi^2 c_s^2}{24R^2}x\rndb{x^2-2R^2-3y^2}\right] \si{.}
\end{align} 
As we are looking at the steady state solutions of the droplet, it can be shown that there are no net translational forces generated by the droplet, 
\begin{align}
 \vectl{F}^{(1)} &= \oint \sqb{\tenss{\sigma} - \rndb{P - P_{ext} - \frac{2\gamma}{R}}\hat{\mathbb{I}}} \cdot \rm{d}\vectl{s}\rndb{\theta}\,\bigg|_{r=R} \nonumber \\
\label{Ftotxy} &= 0 \si{.}
\end{align}
eq. \eqref{Fbbc} ensures that there are no net radial forces from the droplet, and it can be shown similarly that there is no net torque. Therefore, we find that there are no net forces produced by the droplet, as expected at low Reynolds' number. However, it can be shown that there is spatial separation of the equal and opposite forces at the droplet boundary. We show this by taking successive moments of the force at the boundary. As for the force monopole (eq. \eqref{Ftotxy}), increasing moments of the force can be expressed by tensors of increasing order, the dipole and quadrupole moments form the following second and third order tensors respectively:
\begin{align}
\label{dipole} F_{ij}^{(2)} &= \int \limits_0^{2\pi} f_ir_j\si{d}\theta \si{,}\\
\label{quadrupole}  F_{ijk}^{(3)} &= \int \limits_0^{2\pi} f_ir_jr_k\si{d}\theta
\end{align}
where $f_i = -\sqb{\sigma_{im} - \rndb{P - P_{ext} - 2\gamma/R}\delta_{im}}r_m$ evaluated at $r=R$ and the vector $\vectl{r} = (x,y)$. { A detailed calculation of the various force moments is given in appendix \ref{app:moments}, here we just present the results.} The dipole moment eq. \eqref{dipole} gives:
\begin{align}
\label{fij} &F_{ij}^{(2)} = \alpha\rndb{{\delta_{ix}\delta_{jx} - \delta_{iy}\delta_{jy}}}\\
&\si{where} \quad \alpha = \frac{\actt \pi R^2}{2}\rndb{1-\frac{(\pi c_s)^2}{8}} { - \frac{K\pi^3 c_s^2}{8}} \si{.} \nonumber
\end{align}
The coefficient $\alpha$ is always negative and hence the droplet is contractile along the $x$-axis and equally extensile along the $y$-axis. This force dipole is due to the alignment and contraction of the filaments along the $x$-axis, and this can be shown by taking the limit of zero splay, $c_s \rightarrow 0$, where all of the filaments in the droplet are completely aligned. In this limit the dipole moment is maximised, and this explains the behaviour observed in the lattice Boltzmann simulations of an active droplet that is below the threshold concentration of activity to break symmetry. In those simulations, the filaments remain approximately aligned in one direction and the droplet squeezes itself, shortening in the direction of alignment and extending in the perpendicular direction (fig. \ref{fig:vsim}(a)) \cite{Tjhung2012}.\\

\begin{figure*}
\centering
\includegraphics[width=\textwidth]{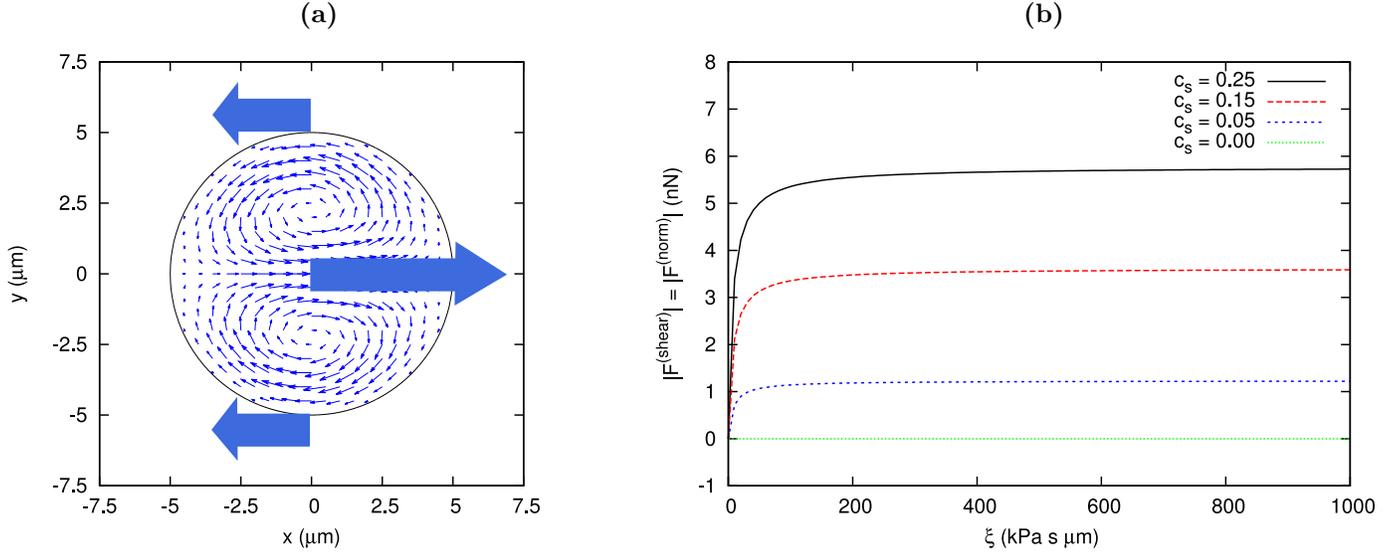}
\caption{\textbf{(a)} Sketch of the droplet motility mechanism in 2D. The large arrows represent the spatially separated forces exerted by the active droplet on a surrounding medium, where the central arrow is the sum of the normal forces in the $x$-direction and the arrows at the top and bottom each contribute half of the total shear force in the $x$-direction. The smaller arrows show the flow profile of the motile droplet in the case of a solid non-slip boundary. \textbf{(b)} Magnitude of the shear and normal forces (which are equal and opposite) plotted against the effective friction coefficient $\xi$ for several values of the splay parameter $c_s$. Note that in the no-splay case these forces are zero, as is the quadrupole moment. Parameter values used are $R=5\mcrm$, $\eta = 10 \si{kPa}\si{s}$ and $\actt = -1 \si{kPa}$.}
\label{fig:forceplots}
\end{figure*}

For a circular droplet at low Reynolds' number, a non-zero force dipole distribution is insufficient for motion \cite{Yoshinaga2013} and in this case it does not vanish in the symmetric limit $c_s \rightarrow 0$. However, the quadrupole moment is directly dependent on the symmetry breaking,
\begin{align}
\label{fijk} &F_{ijk}^{(3)} = \beta\rndb{{-\delta_{ix}\delta_{jx}\delta_{kx} + \delta_{ix}\delta_{jy}\delta_{ky}}}\\
&\si{where} \quad \beta =  \frac{\actt \pi^2 R^3 c_s}{4} {\, + \,O\rndb{c_s^3}} \si{.} \nonumber
\end{align}
Unlike the dipole moment, the forces that make up the quadrupole moment only act in the $x$-direction (shown by the non-zero components of the quadrupole tensor both having $i=x$). It shows that the net normal forces at the front and back of the droplet act in the positive $x$ direction (as $\beta < 0$)  and the shear forces at the sides are equal and opposite, as sketched in fig. \ref{fig:forceplots}(a). In the no-splay limit, $c_s \rightarrow 0$, the quadrupole moment disappears along with the asymmetry in $x$.\\

The quadrupole moment characterises the motility mechanism because of the force distribution it indicates. The shear forces at the sides of the droplet arise from the friction between the fluid and the external medium and hence act to propel the droplet forwards (or the medium rearwards in the droplet rest frame). The normal forces however act to deform the front and back of the droplet asymmetrically in the positive $x$-direction. The magnitude of these shear and normal forces can be calculated by integrating the corresponding elements of the total stress tensor at the boundary,
\begin{align}
\label{shear} F_i^{(\mathrm{shear})} &= -\int_0^{2\pi} \sigma_{ij}r_j \si{d}\theta \,\bigg|_{r=R} \quad \mathrm{where} \; i \neq j \si{,}\\
\label{normal}  F_i^{(\mathrm{norm})} &= -\int_0^{2\pi} \sqb{\sigma_{ii} - \rndb{P - P_{ext} - 2\gamma/R}}r_{i} \si{d}\theta \,\bigg|_{r=R}  \si{.}
\end{align}
These forces only act in the $x$ direction and, in the infinite friction limit, they are given by 
\begin{align}
\label{fmax} F_x^{(\mathrm{shear})} = -F_x^{(\mathrm{norm})} = \frac{\actt \pi^2 R c_s}{2} { \, + \, O\rndb{c_s^3}} \si{.}
\end{align}
The friction dependence of these forces is plotted in fig. \ref{fig:forceplots}(b), and shows that the magnitude of the forces plateaus at large friction, with the maximum at the infinite friction limit given by eq. \eqref{fmax}.\\

We can estimate the migration speed of the droplet through a viscous medium by calculating the average velocity at the boundary. Since we work in the droplet reference frame, if we assume that the velocity of the external fluid will be the same as the internal fluid at $r=R$, then the steady swimming speed is just given by the negative of the average cortical velocity (as shown in \cite{Stone1996}):
\begin{align}
\vectl{v}_{\mathrm{mig}} &= -\frac{1}{2\pi}\int^{2\pi}_{0} \vectl{v}(R,\theta)\si{d}\theta \nonumber\\
\label{vmig} &= \sqb{-\frac{\actt \pi R c_s}{8(\xi R + 2 \eta)} { \, + \, O(c_s^3)}} \hat{\vectl{x}} \si{.}
\end{align}
In this case the effective friction coefficient $\xi$ relates directly to the viscosity of the external fluid, and is inversely proportional to the migration speed. Relating this to the graph in fig. \ref{fig:forceplots}(b) we see that for a very viscous medium, the droplet still exerts almost the maximum amount of force onto the external fluid but the droplet is barely able to move. Conversely, in a low viscosity medium the droplet velocity is high but relatively little force is exerted on the external fluid.\\

To summarise, the dipole moment shows that the droplet behaves like a \emph{puller} (contractile along axis of motion) \cite{Ramaswamy2010}, however this is not sufficient for motility in this case. It is the quadrupole moment that characterises the motility mechanism and symmetry breaking in our system, and the resulting picture resembles the motility mechanism of a \emph{squirmer} (a particle propelled by coordinated beating of cilia on its surface) \cite{Blake1971}. 

{
\subsection{Extension to three dimensions - spherical droplet}
\label{sec:sphere}

{ In ref. \cite{Tjhung2012}, it is found through lattice Boltzmann simulations of the dynamic equations that in 3D, as in 2D, above a certain activity threshold the system stabilises to a splayed steady motile state. However, this motile state becomes unstable at a second higher activity threshold, above which a non-motile aster configuration becomes stable. Therefore, to model the system in 3D, we consider that the activity is between these two thresholds.}\\

By applying appropriate symmetry rules, the calculation for the motile splayed state can be easily extended to 3D for the case of a spherical droplet with radius $R$ and centre at $x=0$. If we assume that the polarisation is still splayed around the $x$-axis uniformly and satisfies $\norm{\vectl{p}} = 1$ then we can define it in cartesian coordinates as
\begin{align}
\vectl{p} =& \sqb{\frac{1}{\sqrt{2}}\sqrt{1 + \cos\rndb{\frac{\pi y}{l}}\cos\rndb{\frac{\pi z}{l}}}\si{,}\, \sin\rndb{\frac{\pi y}{2l}}\cos\rndb{\frac{\pi z}{2l}}\si{,} \right. \notag \\
\label{3dp} &\left. \cos\rndb{\frac{\pi y}{2l}}\sin\rndb{\frac{\pi z}{2l}}} \si{.}
\end{align}
Using the same method as the 2D case we can solve the 3D equations of motion, as shown in appendix \ref{app:sphere}. The resulting flows are analogous to the 2D case, and the resulting force dipole and quadrupole tensors are,
{
\begin{align}
 \label{fij3d} F_{ij}^{(2)} = \tilde{\alpha} \rndb{2\delta_{ix}\delta_{jx} - \delta_{iy}\delta_{jy} - \delta_{iz}\delta_{jz}}
\end{align}
and
\begin{align}
\label{fijk3d} F_{ijk}^{(2)} = \tilde{\beta}\rndb{-2\delta_{ix}\delta_{jx}\delta_{kx} + \delta_{ix}\delta_{jy}\delta_{ky} + \delta_{ix}\delta_{jz}\delta_{kz}} \si{,}
\end{align}
where
\begin{align}
\tilde{\alpha} = \frac{4 \acttp \pi R^3}{9}\rndb{1 - \frac{3\pi^2c_s^2}{20}} { \, - \, \frac{4K\pi^3R\rndb{2+5\nu}}{27}c_s^2} \si{,} \nonumber
\end{align}
and
\begin{align}
\tilde{\beta} = \frac{4 \acttp \pi^2 R^4 c_s}{15} +  O\rndb{c_s^3} \si{,} \nonumber 
\end{align}}
which is calculated from the solutions in equations \eqref{vx3dns}, \eqref{vy3dns}, and \eqref{vz3dns} for the case of $\xi \rightarrow \infty$, { where $\acttp = \act + K\pi^2\rndb{8\nu-1}/(6l^2)$.} As in the 2D circular droplet case, one can see that the dipole moment in eq. \eqref{fij3d} shows that there is a net contraction along the $x$-axis, which is balanced by equal extensile moments in the $y$ and $z$ directions. The second moment shows that the splay induces a preferred direction of motion and the formation of a non-zero quadrupole moment. This shows that there is a net normal force generated at the front and back of the droplet, and an equal net rearward force around the equatorial plane ($yz$ plane) of the droplet.}

\section{Conclusions and further remarks}
\label{sec:conclusion}

The model we have presented here demonstrates analytically how self-propelled motion can be generated in a finite active polar droplet, purely by the internal circulation of material driven by active contractile stresses. By imposing an asymmetric splayed polarisation on the droplet, we have analytically calculated the hydrodynamic steady state of the system. The resulting internal fluid flow coupled with a viscous interaction at the boundary causes a non-zero force quadrupole and leads to self-propulsion of the droplet.\\

In addition, our model predicts flow in the droplet even when the filaments are completely aligned and the droplet is not motile, which is also seen in simulations. In this limit the filament alignment results in a symmetric force dipole moment, which is responsible for the `squashing' of the droplet that is seen in simulations, prior to the development of splay instabilities \cite{Tjhung2012}, {but does not result in overall motility of the droplet.}\\

Our model shows that according to the friction at the boundary and the amount of splay, the intra-droplet flow can undergo transitions between several different patterns. For infinite friction, we observe a transition from zero flow in the no-splay limit to 2 directed vortices for finite splay. In the finite friction case, we observe a transition from 4 symmetric vortices in the no-splay limit to 4 asymmetric vortices for finite splay. This means that for finite splay there is a transition from 4 asymmetric vortices to 2 vortices as the friction goes to infinity. The 4 symmetric vortices are due to the interaction between the active gel and the boundary and consequently the flow is zero in the infinite friction limit due to the boundary condition $v_{\theta}(R,\theta)=0$. The 2 directed vortices are generated by the active stress due to the splay. The 4 asymmetric vortices are a combination of these two effects.\\

Our results predict and explain both the non-motile and motile states observed in lattice Boltzmann simulations of active contractile droplets \cite{Tjhung2012}. Importantly, as we consider a confined active fluid and control the flow field at the boundary, our calculations can extend those of \cite{Tjhung2012} to the case of a more general external medium that can be applied to more biologically relevant cases.\\

Previous theoretical studies using a coarse grained hydrodynamic approach on active gels in bulk and confinement have predicted individual vortices in the flow due to { rotationally invariant filament polarisations \cite{Kruse2004,Woodhouse2012}. Similar aster and vortex type defects in the filament orientation have been observed to self-organise \cite{Aranson2006}} and rotate \cite{Nedelec2002,Head2011} in discrete microscopic models and also experimentally for microtubule filaments \cite{Nedelec1997,Surrey2001}. An important distinction of our model is that the pairs of directed vortices we predict only form in confinement. Experimentally probing {\it in vitro} active gels in confinement is now becoming feasible due to new techniques. Recent examples include confinement of active microtubule-kinesin networks inside microchambers \cite{Laan2012} to measure the forces exerted in these systems and inside fluid droplets in emulsions of water and oil using microfluidic devices \cite{Sanchez2012}.\\ 

The results presented here are complementary to the case of a spherical cell migrating due to the flow of an active gel cortical layer \cite{Hawkins2011}. In that case, a similar force distribution is generated to propel the droplet forwards, only the asymmetry is driven by activity concentration rather than filament polarisation. In this paper we have not explicitly enforced a cortical layer of active gel, but rather we have modelled a complete droplet of active gel, with similar findings. Therefore, it is likely that coupling aspects from both of these simplified systems will lead to greater understanding of friction dependent cell migration in confinement.\\

This analytical study explains how motility can occur in a \emph{in-vitro} active droplet of contractile actomyosin when confined in a 3D environment. Experimental studies of this nature in the future will be important in understanding actomyosin dynamics in cells and also how real active gels differ from the linear model used in these calculations. The motion that we predict is heavily friction dependent, but does not rely on the droplet being adhered to the surrounding medium. In this way the work is also relevant to the study of migration mechanisms in environments where cells are unable to form adhesion complexes that connect the cytoskeleton to the external medium. 

\section*{Acknowledgements}
{We acknowledge the EPSRC for funding, grant reference EP-K503149-1.}

\bibliographystyle{epj}
\bibliography{references}    

\begin{thebibliography}{46}

\bibitem{Ramaswamy2010}
S.~Ramaswamy, Annual Review of Condensed Matter Physics \textbf{1}, 323 (2010)

\bibitem{Toner1998}
J.~Toner, Y.~Tu, Physical Review E \textbf{58}(4), 4828 (1998)

\bibitem{Narayan2007}
V.~Narayan, S.~Ramaswamy, N.~Menon, Science \textbf{317}(5834), 105 (2007),
  ISSN 0036-8075

\bibitem{Kruse2000}
K.~Kruse, F.~J{\"u}licher, Phys. Rev. Lett. \textbf{85}(8), 1778 (2000)

\bibitem{Kruse2004}
K.~Kruse, J.F. Joanny, F.~J{\"{u}}licher, J.~Prost, K.~Sekimoto, Phys. Rev.
  Lett. \textbf{92}(7), 078101 (2004)

\bibitem{Kruse2005}
K.~Kruse, J.F. Joanny, F.~J{\"{u}}licher, J.~Prost, K.~Sekimoto, The European
  Physical Journal E \textbf{16}(1), 5 (2005)

\bibitem{Joanny2009}
J.F. Joanny, J.~Prost, HFSP Journal \textbf{3}(2), 94 (2009)

\bibitem{Mizuno2007}
D.~Mizuno, C.~Tardin, C.~Schmidt, F.~MacKintosh, Science \textbf{315}(5810),
  370 (2007), ISSN 0036-8075

\bibitem{Mackintosh2010}
F.C. Mackintosh, C.F. Schmidt, Curr. Opin. Cell Biol. \textbf{22}(1), 29 (2010)

\bibitem{Silva2011}
M.~Soares~e Silva, M.~Depken, B.~Stuhrmann, M.~Korsten, F.C. MacKintosh, G.H.
  Koenderink, Proceedings of the National Academy of Sciences \textbf{108}(23),
  9408 (2011), ISSN 0027-8424

\bibitem{Kohler2011}
S.~K{\"{o}}hler, V.~Schaller, A.R. Bausch, PloS one \textbf{6}(8), e23798
  (2011), ISSN 1932-6203

\bibitem{Kohler2012}
S.~K{\"{o}}hler, A.R. Bausch, PloS one \textbf{7}(7), e39869 (2012), ISSN
  1932-6203

\bibitem{Voituriez2005}
R.~Voituriez, J.F. Joanny, J.~Prost, Europhys. Lett. \textbf{70}(3), 404 (2005)

\bibitem{Charras2008}
G.~Charras, E.~Paluch, Nat. Rev. Mol. Cell Biol. \textbf{9}(9), 730 (2008),
  ISSN 1471-0072

\bibitem{Hawkins2009}
R.J. Hawkins, M.~Piel, G.~Faure-Andre, A.M. Lennon-Dumenil, J.F. Joanny,
  J.~Prost, R.~Voituriez, Phys. Rev. Lett. \textbf{102}(5), 058103 (2009)

\bibitem{Konstantopoulos2013}
K.~Konstantopoulos, P.H. Wu, D.~Wirtz, Biophys. J. \textbf{104}(2), 279 (2013),
  \texttt{http://dx.doi.org/10.1016/j.bpj.2012.12.016}

\bibitem{Poincloux2011}
R.~Poincloux, O.~Collin, F.~Liz{\'{a}}rraga, M.~Romao, M.~Debray, M.~Piel,
  P.~Chavrier, Proceedings of the National Academy of Sciences \textbf{108}(5),
  1943 (2011)

\bibitem{Weber2013}
M.~Weber, R.~Hauschild, J.~Schwarz, C.~Moussion, I.~de~Vries, D.F. Legler, S.A.
  Luther, T.~Bollenbach, M.~Sixt, Science \textbf{339}(6117), 328 (2013), ISSN
  0036-8075

\bibitem{Tjhung2012}
E.~Tjhung, D.~Marenduzzo, M.E. Cates, Proceedings of the National Academy of
  Sciences \textbf{109}(31), 12381 (2012)

\bibitem{Blanch-Mercader2013}
C.~Blanch-Mercader, J.~Casademunt, Physical Review Letters \textbf{110}(7),
  078102 (2013)

\bibitem{Shao2010}
D.~Shao, W.J. Rappel, H.~Levine, Phys. Rev. Lett. \textbf{105}(10), 108104
  (2010)

\bibitem{Ziebert2011}
F.~Ziebert, S.~Swaminathan, I.S. Aranson, Journal of the Royal Society
  Interface \textbf{9}(70), 1084 (2011)

\bibitem{Sanchez2012}
T.~Sanchez, D.T.N. Chen, S.J. DeCamp, M.~Heymann, Z.~Dogic, Nature
  \textbf{491}, 431 (2012)

\bibitem{Coppin1992}
C.M. Coppin, P.C. Leavis, Biophysical journal \textbf{63}(3), 794 (1992), ISSN
  0006-3495

\bibitem{Gennes1993}
P.G. de~Gennes, J.~Prost, \emph{The Physics of Liquid Crystals}, 2nd~edn.
  (Clarendon Press, Oxford, 1993)

\bibitem{Pleiner1989}
H.~Pleiner, H.~Brand, EPL (Europhysics Letters) \textbf{9}(3), 243 (1989), ISSN
  0295-5075

\bibitem{Furthauer2012}
S.~F{\"{u}}rthauer, M.~Neef, S.~Grill, K.~Kruse, F.~J{\"{u}}licher, New Journal
  of Physics \textbf{14}(2), 023001 (2012), ISSN 1367-2630

\bibitem{Bendix2008}
P.M. Bendix, G.H. Koenderink, D.~Cuvelier, Z.~Dogic, B.N. Koeleman, W.M.
  Brieher, C.M. Field, L.~Mahadevan, D.A. Weitz, Biophys. J. \textbf{94}(8),
  3126 (2008), ISSN 0006-3495

\bibitem{Voituriez2006}
R.~Voituriez, J.F. Joanny, J.~Prost, Phys. Rev. Lett. \textbf{96}(2), 028102
  (2006)

\bibitem{Hawkins2011}
R.J. Hawkins, R.~Poincloux, O.~B{\'{e}}nichou, M.~Piel, P.~Chavrier,
  R.~Voituriez, Biophys. J. \textbf{101}(5), 1041 (2011)

\bibitem{Alberts2008}
B.~Alberts, A.~Johnson, J.~Lewis, M.~Raff, K.~Roberts, P.~Walter,
  \emph{Molecular Biology of the Cell}, 5th~edn. (Garland Science, New York,
  2008)

\bibitem{Kruse2006}
K.~Kruse, J.F. Joanny, F.~J{\"{u}}licher, J.~Prost, Phys. Biol. \textbf{3}(2),
  130 (2006)

\bibitem{Wottawah2005}
F.~Wottawah, S.~Schinkinger, B.~Lincoln, R.~Ananthakrishnan, M.~Romeyke,
  J.~Guck, J.~KÃ¤s, Phys. Rev. Lett. \textbf{94}(9), 098103 (2005), ISSN
  1079-7114

\bibitem{Thoumine1999}
O.~Thoumine, O.~Cardoso, J.J. Meister, Euro. Biophys. J. \textbf{28}(3), 222
  (1999), ISSN 0175-7571, \texttt{http://dx.doi.org/10.1007/s002490050203}

\bibitem{Dai1999}
J.~Dai, M.P. Sheetz, Biophys. J. \textbf{77}(6), 3363 (1999), ISSN 0006-3495

\bibitem{Yoshinaga2013}
N.~Yoshinaga, arXiv preprint arXiv:1307.3120 pp.~-- (2013)

\bibitem{Stone1996}
H.A. Stone, A.D. Samuel, Phys. Rev. Lett. \textbf{77}(19), 4102 (1996)

\bibitem{Blake1971}
J.~Blake, J. Fluid Mech \textbf{46}, 199 (1971)

\bibitem{Woodhouse2012}
F.G. Woodhouse, R.E. Goldstein, Physical review letters \textbf{109}(16),
  168105 (2012)

\bibitem{Aranson2006}
I.S. Aranson, L.S. Tsimring, Physical Review E \textbf{74}(3), 031915 (2006)

\bibitem{Nedelec2002}
F.~N\'{e}d\'{e}lec, The Journal of cell biology \textbf{158}(6), 1005 (2002),
  ISSN 0021-9525

\bibitem{Head2011}
D.A. Head, G.~Gompper, W.J. Briels, Soft Matter \textbf{7}(7), 3116 (2011),
  ISSN 1744-683X, \texttt{http://dx.doi.org/10.1039/C0SM00888E}

\bibitem{Nedelec1997}
F.J. N{\'{e}}d{\'{e}}lec, T.~Surrey, A.C. Maggs, S.~Leibler, Nature
  \textbf{389}(6648), 305 (1997), ISSN 0028-0836,
  \texttt{http://dx.doi.org/10.1038/38532}

\bibitem{Surrey2001}
T.~Surrey, F.~N{\'{e}}d{\'{e}}lec, S.~Leibler, E.~Karsenti, Science
  \textbf{292}(5519), 1167 (2001)

\bibitem{Laan2012}
L.~Laan, N.~Pavin, J.~Husson, G.~Romet-Lemonne, M.~van Duijn, M.P. L\'{o}pez,
  R.D. Vale, F.~J\"{u}licher, S.L. Reck-Peterson, M.~Dogterom, Cell
  \textbf{148}(3), 502 (2012), ISSN 0092-8674

\bibitem{maple}
\emph{Maple-16} (2012), {Maplesoft}, a division of Waterloo Maple
  Inc.,Waterloo, Ontario, Canada

\end{thebibliography}


\numberwithin{equation}{section}
\numberwithin{figure}{section}
\appendix
\section*{Appendices}

\section{Governing equations for hybrid lattice Boltzmann simulations of an active droplet}
\label{app:sims}

The lattice Boltzmann simulations, used to obtain fig. \ref{fig:vsim}, are adapted from \cite{Tjhung2012}. This section summarises the governing equations in these simulations.\\

Firstly, the free energy functional, which governs the passive state of the system, in summation notation is as follows:
\begin{align}
F\sqb{\phi,p_\alpha} &= \int d^{3}r\,\curlb{\frac{a}{4\phi_{cr}^{4}}\phi^{2}(\phi-\phi_{0})^{2} + \frac{k_{\phi}}{2}\left|\nabla \phi \right|^{2}\right. \notag \\ 
\label{Fsim} &\left. - \, \frac{\alpha}{2}(\phi-\phi_{cr}) p_{\alpha}p_{\alpha} + \frac{\alpha}{4}\rndb{p_{\alpha}p_{\alpha}}^2 + \frac{K}{2}\rndb{\partial_\alpha p_\beta}^2} \si{.}
\end{align}
In this model, $\phi$ is the activity concentration, and so $\phi=0$ in the passive phase and $\phi > \phi_{cr}$ in the active phase. The first term in eq. \eqref{Fsim}, with coefficient $a$, gives free energy minima for $\phi$ at $\phi = 0$ and $\phi = \phi_{0} > \phi_{cr}$. The coefficient $k_{\phi}$ contributes to the interfacial tension, and $\alpha$ characterises the isotropic to nematic transition (this term couples $\norm{\vectl{p}} = 0$ to the passive phase and $\norm{\vectl{p}} = 1$ to the active phase). The final term is the distortion free energy from eq. \eqref{freeE} where $K$ is the elastic constant in the one constant approximation $K_1=K_2=K_3=K$ and we assume no specific anchoring at the boundary.\\

The total activity in the system is conserved, so the time evolution of $\phi$ is calculated using a convective-diffusion equation at each time step:
\begin{align}
\label{dphidt} \pdiff{\phi}{t} + (v_\alpha\partial_\alpha)\phi = M\partial_\alpha^2\frac{\delta F}{\delta \phi} \si{,}
\end{align}
where $M$ is related to the diffusion. The polarisation dynamics are then governed by eq.\eqref{dpdt} with $\lambda = 0$, as this term only has an effect when $\norm{p} \neq 1$ in the active phase.\\

Lattice Boltzmann techniques are used to satisfy the incompressible Navier-Stokes equations (the time dependent version of eqs. \eqref{incgen} and \eqref{fbgen}),
\begin{align}
\label{incsim} \partial_\alpha v_\alpha  &=  0\\
\label{fullns} \rho\rndb{\pdiff{}{t}+v_k\partial_k }v_\alpha  &= \partial_\beta\rndb{\sigma_{\beta\alpha}^{\mathrm{tot}} - P\delta_{\alpha \beta}}  \si{.}
\end{align}
The tensor $\tenss{\sigma}^{\mathrm{tot}}$ is the stress tensor of eq. \eqref{totstress} with the addition of the interfacial stress given by,
\begin{align}
\label{intstress} \sigma_{\alpha \beta}^{\mathrm{interface}} = - \phi \frac{\delta F}{\delta \phi}\delta_{\alpha \beta} - \pdiff{f}{(\partial_\beta\phi)}\partial_\alpha \phi \si{,}
\end{align}
where $f$ is the free energy density in eq.\eqref{Fsim}.\\

We initialise the simulation by assuming an aligned polarisation field $\vectl{p} = \hat{\vectl{x}}$ inside a circular droplet in the active phase. The initial radius of the droplet is set to approximately $1/5$ of the total grid size so that the drop is isolated and the boundaries are periodic to allow the droplet to migrate freely.

\section{Calculation of steady state for circular droplet}
\label{app:sols}

The assumed power series solutions in eqs. \eqref{vxpow}, \eqref{vypow}, and \eqref{Ppow} can be reduced by assuming that they will be symmetric about the $x$-axis, since the system, defined by the polarisation and the boundary conditions, obeys this symmetry. This implies that $a_{m,2n+1} = b_{m,2n} = c_{m,2n+1} = 0$ and leaves,
\begin{align}
\label{vxpows} v_x(x,y) &= \sum\limits_{n=0}^{\infty}\sum\limits_{m=0}^{\infty}{a_{m,2n}x^my^{2n}}  \mathrm{,}\\
\label{vypows} v_y(x,y) &= \sum\limits_{n=0}^{\infty}\sum\limits_{m=0}^{\infty}{b_{m,2n+1}x^my^{2n+1}}  \mathrm{,}\\
\label{Ppows} P(x,y) &= \sum\limits_{n=0}^{\infty}\sum\limits_{m=0}^{\infty}{c_{m,2n}x^my^{2n}}  \mathrm{.}
\end{align}
We then substitute these into the governing partial differential equations and boundary conditions; eqs. \eqref{pde3}, \eqref{pde1s}, \eqref{pde2s}, \eqref{circbc1}, and \eqref{circbc2} and solve these simultaneously to determine the values of all the constants $a_{m,n}$, $b_{m,n}$, and $c_{m,n}$.\\

The incompressibility condition of eq. \eqref{pde3} becomes,
\begin{align}
\sum\limits_{n=0}^{\infty}\sum\limits_{m=0}^{\infty} \sqb{a_{m+1,2n}(m+1) + b_{m,2n+1}(2n+1)}x^m y^{2n} = 0 \si{,}
\end{align}
which, comparing coefficients, gives the following set of equations for all values of $n$ and $m$,
\begin{align}
\label{pde3a} b_{m,2n+1} = -\frac{m+1}{2n+1}a_{m+1,2n} \rm.
\end{align}\\

The $x$-component of the force balance eq. \eqref{pde1s} becomes:
\begin{align}
&\sum\limits_{n=0}^{\infty}\sum\limits_{m=0}^{\infty}x^m y^{2n}\left[\eta a_{m+2,2n}\rndb{m+2}\rndb{m+1} \right. \notag \\
\vphantom{\sum\limits_{n=0}^{\infty}} &\left.+ \eta a_{m,2n+2}\rndb{2n+2}\rndb{2n+1} - c_{m+1,2n}\rndb{m+1} \right]  \notag \\
\label{pde1sum}  &= \frac{\actt\pi c_s}{2R}\sqb{1-\frac{\pi^2 y^2 c_s^2}{2R^2}} \mathrm{.}
\end{align}  
This leads to the following set of simultaneous equations,\\
($m=n=0$):
\begin{align}
\label{pde1a} c_{1,0} = 2\eta \rndb{a_{2,0} + a_{0,2}} - \frac{\actt\pi c_s}{2R} \si{,}
\end{align}
($m=0$, $n=1$):
\begin{align}
\label{pde1b} c_{1,2} = 2\eta \rndb{a_{2,2} + 6a_{0,4}} + \frac{\actt\pi^3 c_s^3}{4R^3} \si{,}
\end{align}
(all other $m$ and $n$ combinations):
\begin{align}
c_{m+1,2n} = \frac{\eta}{m+1}\left[a_{m+2,2n}\rndb{m+1}\rndb{m+2}\right. \notag\\
\label{pde1c} \vphantom{\frac{1}{2}}\left. +\, a_{m,2n+2}\rndb{2n+1}\rndb{2n+2}\right] \si{.}
\end{align}

The $y$-component of the force balance eq. \eqref{pde2s} becomes:
\begin{align}
&\sum\limits_{n=0}^{\infty}\sum\limits_{m=0}^{\infty} x^m y^{2n+1}\left[\eta b_{m+2,2n+1}\rndb{m+2}\rndb{m+1} \right. \notag \\
& \left. + \eta b_{m,2n+3}\rndb{2n+3}\rndb{2n+2} -  c_{m,2n+2}\rndb{2n+2} \right] \notag \\
\label{pde2sum} &= \frac{\actt\pi^2 c_s^2 y}{2R^2} \mathrm{.}
\end{align}
For $m=n=0$, this gives,
\begin{align}
\label{pde2a} c_{0,2} = \eta\rndb{b_{2,1} + 3b_{0,3}} - \frac{\actt \pi^2 c_s^2}{4 R^2} \si{,}
\end{align}
and for all other $m$ and $n$ values,
\begin{align}
c_{m,2n+2} = \frac{\eta}{2n+2}\left[b_{m+2,2n+1}(m+1)(m+2)\right. \notag\\
\label{pde2b} \left. + b_{m,2n+3}(2n+2)(2n+3)\right] \si{.}
\end{align}

To apply the circular boundary conditions, we transform the velocity solutions of eqs. \eqref{vxpows} and \eqref{vypows} into plane polar coordinates. We use binomial expansions on terms of the form $\sin^{2n}\rndb{\theta}$ so that the expressions can be written in terms of powers of $\cos\rndb{\theta}$, as such:
\begin{align}
\label{binom} \sin^{2n}\rndb{\theta} = (1-\cos^2\rndb{\theta})^n = \sum\limits_{k=0}^{n}(-1)^k
\left(\begin{array}{c}
n\\
k
\end{array}\right)
\cos^{2k}\rndb{\theta} \si{.}
\end{align}
 Then, through further manipulation we acquire the following equations for the plane polar components of the velocity,
\begin{align}
v_r &= \, \sum\limits_{j=0}^{\infty}\sum\limits_{k=0}^{\bar{j}}(-1)^k\left\{\sqb{\sum\limits_{n=k}^{\infty}
\left(\begin{array}{c}
n\\
k
\end{array}\right)
a_{j-2k-1,2n}r^{j+2n-2k-1}} \right. \notag \\
\label{vrpows} & \left. + \sqb{\sum\limits_{n=k-1}^{\infty}
\left(\begin{array}{c}
n+1\\
k
\end{array}\right)
b_{j-2k,2n+1}r^{j+2n-2k+1}}\right\}\cos^j\rndb{\theta} \si{,}\\
v_{\theta} &= \, \sin\rndb{\theta}\sum\limits_{j=0}^{\infty}\sum\limits_{k=0}^{\bar{j}}(-1)^k\left[\sum\limits_{n=k}^{\infty}
\left(\begin{array}{c}
n\\
k
\end{array}\right)
\left(-a_{j-2k,2n}  \right. \right. \notag \\
\label{vtpows} & \left. \left. + \, b_{j-2k-1,2n+1}\right)r^{j+2n-2k}\right]\cos^j\rndb{\theta} \si{.}
\end{align}
Here, the substitution $j=m+2k$ was used so that we can compare coefficients of powers of $\cos\rndb{\theta}$, and $\bar{j}$ is just,
\begin{align}
\label{jbar} \bar{j} = 
\begin{cases}
j/2 \quad \quad \quad \;\; \mathrm{if} \;\; j = \mathrm{even} \vspace{2mm} \\ (j-1)/2 \quad \mathrm{if} \;\; j = \mathrm{odd} \si{.}
\end{cases}
\end{align}
Note also that  eqs. \eqref{vrpows} and \eqref{vtpows} contain coefficients with negative subscripts (\emph{e.g.} $b_{0,-1}$), which are defined as equal to $0$, as they are not in the original expansion. They are only included so that $v_r$ and $v_\theta$ can be written in such a general form. Substituting eq. \eqref{vrpows} into the impermeable boundary condition eq. \eqref{circbc1} we acquire the following set of simultaneous equations for all values of $j$:
\begin{align}
&\sum\limits_{k=0}^{\bar{j}}(-1)^k\left\{\sqb{\sum\limits_{n=k}^{\infty}
\left(\begin{array}{c}
n\\
k
\end{array}\right)
a_{j-2k-1,2n}R^{j+2n-2k-1}} \right. \notag \\
\label{bc1s} & \left. + \sqb{\sum\limits_{n=k-1}^{\infty}
\left(\begin{array}{c}
n+1\\
k
\end{array}\right)
b_{j-2k,2n+1}R^{j+2n-2k+1}} \right\} = 0 \si{.}
\end{align}

The final set of simultaneous equations comes from the friction boundary condition of eq. \eqref{circbc2}, which written in full is:
\begin{align}
& \eta r \sqb{\pdiff{(v_{\theta}/r)}{r} + \frac{1}{r^2}\pdiff{v_r}{\theta}}  - \curlb{\actt\sqb{\frac{\pi c_s r}{2R}\rndb{-1 + \frac{\pi^2 c_s^2 r^2}{6R^2}} \right. \right. \notag \\
& + \left(-1 + \frac{\pi^2 c_s^2 r^2}{2R^2}\right)\cos(\theta) + \frac{\pi c_s r}{R}\rndb{1 - \frac{\pi^2 c_s^2 r^2}{4R^2}}\cos^2(\theta) \notag \\ 
& \left.\left.  - \frac{\pi^2 c_s^2 r^2}{2R^2}\cos^3(\theta) +  \frac{\pi^3 c_s^3 r^3}{6 R^3}\cos^4(\theta)} \right. \notag \\
\label{bc2}  & \left. + \, \frac{K\pi^2 c_s^2 r}{4R^2}\cos(\theta)}\sin(\theta)  = -\xi v_{\theta}\,\bigg|_{r=R} \si{.}
\end{align}
Here we have approximated the stress tensor component $\sigma_{r\theta}$ up to third order terms in $r/l$, because this is differentiated in the force balance equation, which we approximated to second order in this calculation (eqs. \eqref{pde1s} and \eqref{pde2s}). Substituting eqs. \eqref{vrpows} and \eqref{vtpows} into eq. \eqref{bc2} and comparing coefficients of $\cos^j(\theta)$ gives the following set of simultaneous equations,\\
($j = 0$)
\begin{align}
\label{bc2j0} \sum\limits_{n=0}^{\infty}\left[\xi+(2n-1)\frac{\eta}{R}\right]a_{0,2n}R^{2n} = \frac{\actt \pi c_s}{2}\left(1 - \frac{\pi^2 c_s^2}{6}\right)\si{,}
\end{align}
($j=1$)
\begin{align}
\sum\limits_{n=0}^{\infty}\rndb{\xi + 2n\frac{\eta}{R}}\rndb{a_{1,2n} - b_{0,2n+1}}R^{2n+1} \notag \\ 
\label{bc2j1} = \actt\rndb{1 - \frac{\pi^2 c_s^2}{2}} - \frac{K\pi^2c_s^2}{4} \si{,}
\end{align}
($j=2$)
\begin{align}
\sum\limits_{k=0}^1 (-1)^k \sum\limits_{n=k}^{\infty}
\left(\begin{array}{c}
n\\
k
\end{array}\right)
\sqb{\xi + \rndb{1+2n-2k}\frac{\eta}{R}}\left(a_{2-2k,2n} \right.\notag\\
\label{bc2j2} \left.-\, b_{1-2k,2n+1}\right)R^{2(1+n-k)} = \actt\pi c_s\rndb{-1+\frac{\pi^2c_s^2}{4}} \si{,}
\end{align}
($j=3$)
\begin{align}
\sum\limits_{k=0}^1 (-1)^k \sum\limits_{n=k}^{\infty}
\left(\begin{array}{c}
n\\
k
\end{array}\right)
\sqb{\xi + \rndb{1+n-k}\frac{2\eta}{R}}\rndb{a_{3-2k,2n} \right. \notag \\
\label{bc2j3} \left. - \,b_{2-2k,2n+1}}R^{3+2n-2k} =-\actt\frac{\pi^2c_s^2}{2} \si{,}
\end{align}
($j=4$)
\begin{align}
\sum\limits_{k=0}^2  (-1)^k \sum\limits_{n=k}^{\infty}
\left(\begin{array}{c}
n\\
k
\end{array}\right)
\sqb{\xi + \rndb{3+2n-2k}\frac{\eta}{R}} \rndb{a_{4-2k,2n} \right. \notag \\
\label{bc2j4} \left. -\, b_{3-2k,2n+1}}R^{2(2+n-k)} = \actt\frac{\pi^3c_s^3}{6}
\end{align}
(all other values of $j$)
\begin{align}
\sum\limits_{k=0}^{\bar{j}}  (-1)^k \sum\limits_{n=k}^{\infty}
\left(\begin{array}{c}
n\\
k
\end{array}\right)
\sqb{\xi + \rndb{j+2n-2k-1}\frac{\eta}{R}} \rndb{a_{j-2k,2n} \right. \notag \\
\label{bc2j5} \left. - b_{j-2k-1,2n+1}}R^{j+2n-2k} \vphantom{\sum\limits^a_b} = 0 \si{.}
\end{align}
We simultaneously solve  eqs. \eqref{pde3a},\eqref{pde1a}, \eqref{pde1b}, \eqref{pde1c}, \eqref{pde2a}, \eqref{pde2b}, \eqref{bc1s}, \eqref{bc2j0}, \eqref{bc2j1}, \eqref{bc2j2}, \eqref{bc2j3}, \eqref{bc2j4} and \eqref{bc2j5} using Maple$^{\rm{TM}}$ \cite{maple} for the finite case where we truncate the general solutions eqs. \eqref{vxpow}, \eqref{vypow} and \eqref{Ppow} by assuming that $a_{m,n} = b_{m,n} = c_{m,n} = 0$ when $m+n \geq i$ where $i$ is a finite integer. By solving at various values of $i$ we find that the general solutions are attained for $i\geq6$ because when coefficients that have  $m + n > 6$ are included these are found to be $0$. This truncation occurs because we have approximated the equations of motion to second order in $r/l$. More accurate solutions can be obtained by approximating these to higher order, and for each extra order included, the minimum value of $i$ required for complete solutions increases by 1. However, to our knowledge, the solutions cannot be generalised analytically to solve for the infinite order case.\\

Finally, the complete analytical solutions are given by eqs. \eqref{vxfull}, \eqref{vyfull}, and \eqref{Pfull}.
\begin{figure*}
\begin{align}
v_{x} =& \actt \curlb{-\frac{\pi R c_s}{8 \rndb{ \xi R+2\eta }} \sqb{1 + \frac{\pi^2c_s^2}{48 \eta}\rndb{\xi R -2 \eta}}  - \frac{x}{2 \rndb{\xi R+4\eta}}\rndb{1 - \frac{\pi^2c_s^2}{4}} \right. \notag \\
& + \, \frac{\pi c_s}{2R\rndb{\xi R + 6\eta}\rndb{\xi R + 2\eta}} \sqb{\xi R x^2 + 3\eta\rndb{x^2+y^2} + \frac{\pi^2 c_s^2}{32} \rndb{  \xi R\rndb{9y^2-x^2} +  6\eta\rndb{y^2 - x^2} + \frac{\xi^2 R^2}{3\eta}\rndb{x^2+3y^2}}} \notag \\
& + \, \frac{x}{2R^2\rndb{\xi R + 4\eta}\rndb{\xi R + 8 \eta}}\sqb{\rndb{\xi R + 8\eta}\rndb{x^2+3y^2} - \frac{\pi^2 c_s^2}{2}\rndb{\xi R x^2 + 6\eta\rndb{x^2 + y^2}}} \notag \\
& + \,\frac{\pi c_s}{8R^3\rndb{\xi R + 6\eta}\rndb{\xi R + 10\eta}}\sqb{\frac{\pi^2 c_s^2}{12}\rndb{\vphantom{\frac{a^b}{c^d}} \xi R \rndb{3x^4 - 30x^2y^2 - 25y^4} + 45\eta\rndb{x^4 - 2x^2y^2 - 3y^4}  \right. \right. \notag \\
& \,  \left. \left. - \frac{\xi^2 R^2}{4\eta}\rndb{x^4 + 6x^2y^2 + 5y^4}} - \rndb{\xi R + 10\eta}\rndb{3x^4 + 6x^2y^2 - 5y^4}} + \frac{\pi^2 c_s^2 x\rndb{x^4 - 5y^4}}{8R^4\rndb{\xi R + 8\eta}} \notag \\
\label{vxfull} & \, \left. -\frac{\pi^3 c_s^3 \rndb{5x^6 - 15x^4y^2 - 45x^2y^4 + 7y^6}}{192 R^5\rndb{\xi R + 10\eta}}} -  \frac{K\pi^2 c_s^2 x\rndb{R^2-x^2-3y^2}}{8R^3(\xi R + 4\eta)}
\end{align}
\begin{align}
v_y =& \actt \curlb{\frac{y}{2\rndb{\xi R + 4\eta}}\rndb{1-\frac{\pi^2 c_s^2}{4}} - \frac{\pi c_s x y}{R\rndb{\xi R + 6\eta}\rndb{\xi R + 2\eta}}\sqb{\xi R + 3\eta + \frac{\pi^2 c_s^2}{32}\rndb{\frac{(\xi R)^2}{3\eta} - \xi R - 6\eta}} \right. \notag \\ 
&  - \, \frac{y}{2R^2\rndb{\xi R + 4\eta}\rndb{\xi R + 8 \eta}}\sqb{\rndb{\xi R + 8\eta}\rndb{3x^2+y^2} - \frac{\pi^2 c_s^2}{2}\rndb{3\xi R x^2 + 2\eta\rndb{9x^2 + y^2}}} \notag \\
& + \frac{\pi c_s x y}{2R^3\rndb{\xi R + 6\eta}\rndb{\xi R + 10\eta}}\sqb{\rndb{\xi R + 10\eta}\rndb{3x^2+y^2} + \frac{\pi^2 c_s^2}{12}\rndb{\frac{\xi^2 R^2}{4\eta}\rndb{x^2+y^2} + \xi R \rndb{5y^2-3x^2}  \right. \right. \notag \\
\label{vyfull} & \left. \left. \left. \vphantom{\frac{a^b}{c^d}} + \, 15\eta\rndb{y^2-3x^2}}} -  \frac{\pi^2 c_s^2 y\rndb{5x^4 - y^4}}{8R^4\rndb{\xi R + 8\eta}} + \frac{\pi^3 c_s^3 y\rndb{15x^5 - 10x^3y^2 - 9xy^4} }{96R^5\rndb{\xi R + 10\eta}}\vphantom{\frac{a^b}{c^d}}}  + \frac{K\pi^2 c_s^2  y\rndb{R^2-3x^2-y^2}}{8R^3(\xi R + 4\eta)}
\end{align}
\begin{align}
P =& \, c_{0,0} + \actt\curlb{- \frac{\xi \pi c_s x}{2\rndb{\xi R + 2 \eta}}\rndb{1-\frac{\pi^2 c_s^2}{12}} + \frac{3 \eta}{R^2\rndb{\xi R + 4 \eta}}\sqb{\rndb{x^2-y^2} - \frac{\pi^2 c_s^2}{12\eta}\rndb{\xi R y^2 + \eta\rndb{3x^2 + y^2}}} \right. \notag \\
& - \, \frac{2\eta\pi c_s x\rndb{x^2 - 3y^2}}{R^3\rndb{\xi R + 6\eta}}\rndb{1+\frac{\pi^2c_s^2}{96\eta}\rndb{\xi R - 6\eta}} + \frac{5\eta\pi^2 c_s^2 \rndb{x^4 - 6x^2y^2 + y^4}}{8R^4\rndb{\xi R + 8\eta}} \notag \\
\label{Pfull} & \left. - \, \frac{\eta \pi^3 c_s^3  x \rndb{x^4 - 10x^2y^2 + 5y^4}}{8R^5\rndb{\xi R + 10\eta}}}  + \frac{3K\eta\pi^2c_s^2 \rndb{x^2-y^2}}{4R^3(\xi R + 4\eta)} \si{.} 
\end{align}
\end{figure*}

\section{Calculation of force dipole and quadrupole moments}
\label{app:moments}

{ Taking the expression for the net force on the droplet boundary from eq. \eqref{Ftotxy}, we can find the force dipole moment with respect to $r_j$ as follows,
\begin{align}
 F_{ij}^{(2)} &= \oint r_j\sqb{\sigma_{im} - \rndb{P - P_{ext} - \frac{2\gamma}{R}}\delta_{im}} \rndb{\rm{d}\boldsymbol{s}\rndb{\theta}}_m\,\bigg|_{r=R} \nonumber \\
\label{Dipolefull} &= 0 \si{.}
\end{align}
Where $\vectl{s}\rndb{\theta}=R\hat{\vects{\theta}}$ and so,
\begin{align}
\rm{d}\vectl{s}\rndb{\theta} = R\diff{\hat{\vects{\theta}}}{\theta}\rm{d}\theta = -R\rm{d}\theta\hat{\vectl{r}}\si{.}
\end{align}
Therefore, the dipole and quadrupole moments can be rewritten as
\begin{align}
\label{dipolea} F_{ij}^{(2)} &= \int \limits_0^{2\pi} f_ir_j\si{d}\theta \si{,}\\
\label{quadrupolea}  F_{ijk}^{(3)} &= \int \limits_0^{2\pi} f_ir_jr_k\si{d}\theta
\end{align}
where $f_i = -\sqb{\sigma_{im} - \rndb{P - P_{ext} - 2\gamma/R}\delta_{im}}r_m$ evaluated at $r=R$. We can find $f_i$ by first calculating the stress components $\sigma_{ij}$:
\begin{align}
\label{sxx} \sigma_{xx} &= 2\eta\partial_xv_x + \frac{K\pi^2 c_s^2}{8R^2} - \actt\cos^2\rndb{\frac{\pi c_s y}{2R}} \si{,}\\
\label{sxy} \sigma_{xy} = \sigma_{yx} &= \eta\rndb{\partial_xv_y +\partial_yv_x} - \frac{\actt}{2}\sin\rndb{\frac{\pi c_s y}{R}} \si{,}\\
\label{syy} \sigma_{yy} &= 2\eta\partial_yv_y - \frac{K\pi^2 c_s^2}{8R^2} - \actt\sin^2\rndb{\frac{\pi c_s y}{2R}} \si{.}
\end{align}
We can then substitute in the flow solutions for the $\xi \rightarrow \infty$ case from eqs. \eqref{vxns} and \eqref{vyns} into eqs. \eqref{sxx}, \eqref{sxy} and \eqref{syy} and then Taylor expand the active terms to third order in $y/l$ (to be consistent with expanding the force balance equations to second order) to get:
\begin{align}
\sigma_{xx} &= \frac{K(\pi c_s)^2}{8R^2} -\frac{\actt}{2}\curlb{1 - \frac{(\pi c_s)^2y^2}{2R^2} \right. \notag \\
\label{sxx1} &\left. \, + \frac{(\pi c_s)^3x}{24R}\sqb{3\rndb{\frac{y}{R}}^2 + \rndb{\frac{x}{R}}^2-1}}  \si{,}\\
\label{sxy1} \sigma_{xy} &= -\frac{\actt \pi c_s y}{2R}\curlb{1 - \frac{(\pi c_s)^2}{24}\sqb{1 + 2\rndb{\frac{y}{R}}^2}} \si{,}\\
\label{syy1} \sigma_{yy} &= -\sigma_{xx} \si{,}
\end{align}
where we have made the substitution $l=R/c_s$. Next, we substitute these stress components from eqs. \eqref{sxx1}, \eqref{sxy1} and \eqref{syy1} along with the pressure from eq. \eqref{Pns} into the expression for $f_i$. Then, converting to polar coordinates ($x=r\cos(\theta)$ and $y=r\sin(\theta)$) and evaluating at $r=R$ we get:
\begin{align}
f_x\bigg|_{r=R} &= \actt R \curlb{-\frac{\cos(\theta)}{2} + (\pi c_s)\sqb{\cos^2(\theta)-2} \right. \notag\\ 
& \left. \,+\, \frac{(\pi c_s)^2}{16}\sqb{7\cos(\theta) - 8\cos^3(\theta)} \right. \notag \\ 
& \left. \,+\, \frac{(\pi c_s)^3}{48}\sqb{3-12\cos^2(\theta) + 8\cos^4(\theta)}} \notag \\
\label{fxa} & \,+\, \frac{K(\pi c_s)^2}{8R}\cos(\theta) \si{,}\\
 \label{fya} f_y\bigg|_{r=R} &= \curlb{\frac{\actt R}{2} \sqb{-1 + \frac{(\pi c_s)^2}{8}} + \frac{K(\pi c_s)^2}{8R}}\sin(\theta) \si{.}
\end{align}
We then use these expressions to compute the various definite integrals given in eqs. \eqref{Ftotxy}, \eqref{fij}, and \eqref{fijk} to get the values for $F_i^{(1)}$, $F_{ij}^{(2)}$, and $F_{ijk}^{(3)}$.}

\section{Overview of calculation in three dimensions}
\label{app:sphere}

\begin{figure*}
\centering
\includegraphics[width=\textwidth]{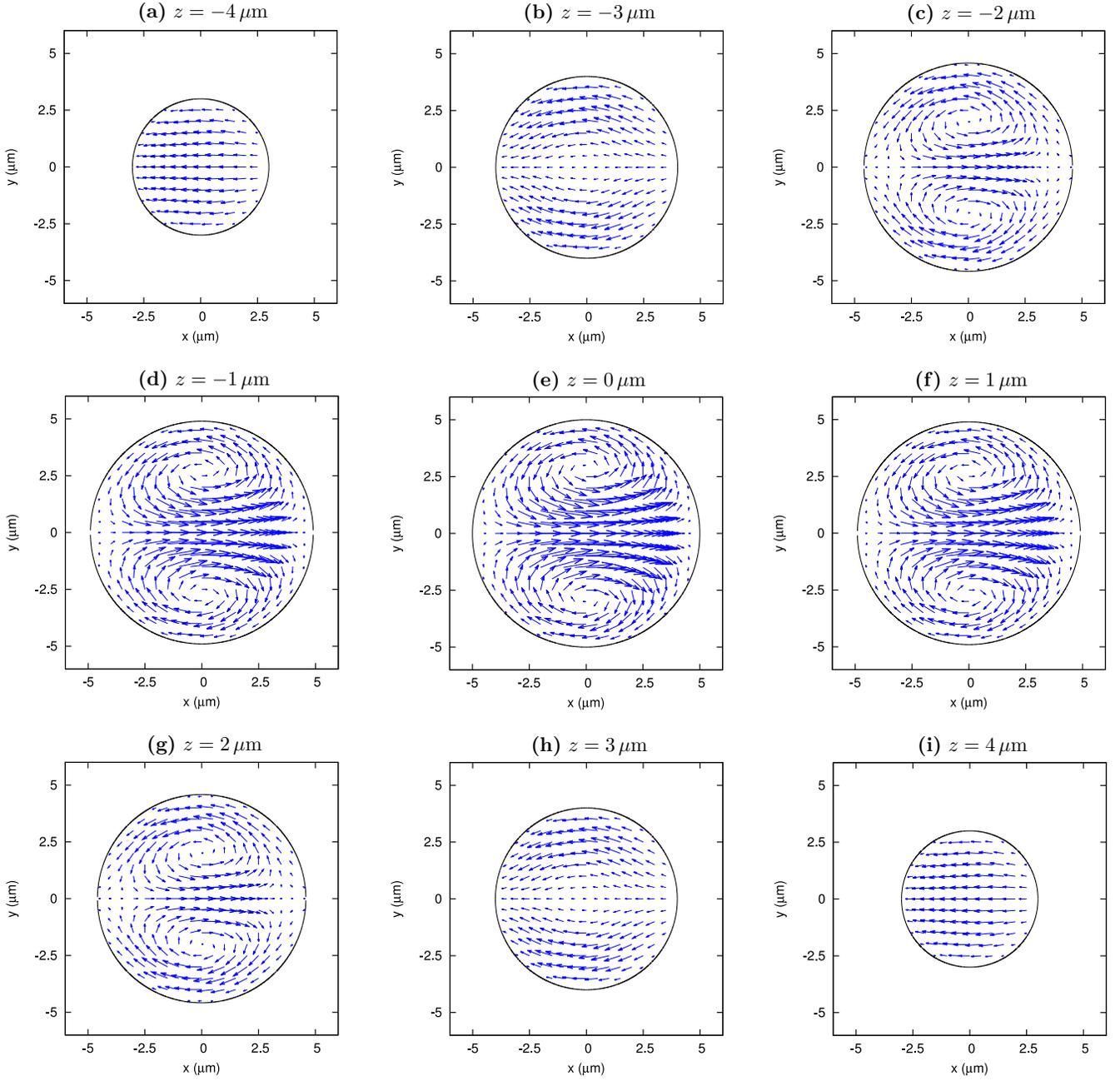}
\label{fig:sphere}
\caption{Vector plots of the $x$ and $y$ components of the velocity profile in a spherical droplet with centre at the origin. Each plot shows a different slice at the labelled $z$ values. The parameter values used are $\xi \rightarrow \infty$, $R = 5 \mcrm$, $c_s = 0.25$, $\acttp = -1 \si{kPa}$ and $\eta = 10 \si{kPa} \si{s}$.}
\end{figure*}

Using the polarisation field from eq. \eqref{3dp}, we find that the governing partial differential equations, to a second order approximation (the 3D analogue of eqs. \eqref{pde1s} and \eqref{pde2s}), become
\begin{align}
 \eta {\nabla}^2v_x &= \frac{\acttp\pi c_s}{R}\sqb{1-\frac{3\pi^2 c_s^2}{8R^2}\rndb{y^2+z^2}}  \notag \\
\label{3dpde1s} & + \, \frac{K\pi^3 c_s^3}{24R^3}\rndb{2\nu-1} + \partial_x P \si{,} \\
\label{3dpde2s} \eta {\nabla}^2v_y &= \frac{3\actb\pi^2 c_s^2}{4R^2}y + \partial_y P \si{,}\\
\label{3dpde3s} \eta {\nabla}^2v_z &= \frac{3\actb\pi^2 c_s^2}{4R^2}z + \partial_z P \si{,}
\end{align}
{ where $\acttp = \act + K\pi^2 c_s^2\rndb{8\nu-1}/(6R^2)$ and $\actb = \act + 2K\nu\pi^2 c_s^2/(3R^2)$}, with the incompressibility condition now defined as
\begin{align}
\label{3dinc}  \partial_x v_x + \partial_y v_y + \partial_z v_z = 0 \si{.}
\end{align}
We can use the boundary conditions of eqs. \eqref{circbc1} and \eqref{circbc2} as they are, but now $r$ is the spherical radial coordinate and $\theta$ is the angle between $r$ and the $z$ axis. We also need an extra boundary condition that defines the friction similarly for the tangential stress in the $\varphi$ direction, where $\varphi$ is the angle between projection of $r$ on to the $xy$-plane and the $x$-axis. This is simply given by the condition $\sigma_{r\varphi} = -\xi v_\varphi$ at $r = R$. Clearly, the $y$ and $z$-directions here are indistinguishable and we apply this to the final solution by ensuring that
\begin{align}
\label{symmyz} v_y \big|_{y=z, z=y} &= v_z  \si{,}\\
\label{symmyzx} v_x \big|_{y=z, z=y} &= v_x  \si{,} \; \si{and}\\
\label{symmyzP} P \big|_{y=z, z=y} &= P \si{.}
\end{align}
As one would expect, the active terms in the solution are just a 3D projection of the 2D case. In the infinite friction limit ($\xi \rightarrow \infty$) the flow solutions are
\begin{align}
\label{vx3dns} v_{x} &= \frac{3\acttp \pi^3 c_s^3}{560 \eta R^3}\rndb{R^2-x^2-y^2-z^2}\rndb{x^2+3y^2+3z^2-R^2} \si{,} \\
\label{vy3dns} v_{y} &= -\frac{3\acttp \pi^3 c_s^3}{280 \eta R^3}xy\rndb{R^2-x^2-y^2-z^2} \si{,} \\
\label{vz3dns} v_{z} &= -\frac{3\acttp \pi^3 c_s^3}{280 \eta R^3}xz\rndb{R^2-x^2-y^2-z^2} \si{.} \\
\end{align} 
Plots of these non-slip solutions are presented in fig. \ref{fig:sphere}.
\end{document}